\documentclass{emulateapj}
\usepackage{times}
\usepackage[]{graphicx,epsfig,rotating}

\begin{document}

\bibliographystyle{apj}

\title{The Environmental Dependence of the Luminosity-Size Relation for Galaxies}
\author{\medskip Preethi B. Nair\footnote{Present address: INAF - Astronomical Observatory of Bologna, Via Ranzani 1, I - 40127 Bologna, ITALY}}
\affil{
   Department of Astronomy \& Astrophysics,
   University of Toronto, 50 St. George Street,
   Toronto, ON, M5S~3H4.
}
\slugcomment{Submitted Nov 5, 2009; Accepted by ApJ April 6, 2010}
\email{preethi.nair@oabo.inaf.it}

\author{\medskip Sidney van den Bergh }
\affil{Dominion Astrophysical Observatory,
Herzberg Institute of Astrophysics,
National Research Council of Canada,
5071 West Saanich Road,
Victoria, British Columbia,
Canada   V9E 2E7
}
\email{sidney.vandenbergh@nrc-cnrc.gov.ca}

\author{Roberto G. Abraham }
\affil{
   Department of Astronomy \& Astrophysics,
   University of Toronto, 50 St. George Street,
   Toronto, ON, M5S~3H4.
}
\email{abraham@astro.utoronto.ca}

\begin{abstract}
We have examined the luminosity-size relationship as a function of environment for 12150 SDSS galaxies with precise visual 
classifications from the catalog of \cite{Nair:2010p22790}. Our analysis is subdivided into investigations of early-type
galaxies and late-type galaxies. Early-type galaxies reveal a surprisingly tight luminosity-size relation. The dispersion in luminosity about the fiducial relation is only $\sim 0.14$ dex (0.35 mag), even though the sample contains galaxies which differ by a 
factor of almost $100$ in luminosity. The dispersion about the luminosity-size relation is comparable to the dispersion 
about the fundamental plane, even though the luminosity-size relation is fundamentally simpler and computed using purely photometric parameters. The key contributors to the dispersion about the luminosity-size relation are found
to be color and central concentration.
Expanding our analysis to the full range of morphological types, we show that the slope, 
zero point, and scatter about the luminosity-size relation is independent of environmental density. 
Our study thus indicates that whatever process is building galaxies is doing so in a way that preserves fundamental
scaling laws even as the typical luminosity of galaxies changes with environment. However, the distribution
of galaxies {\em along} the luminosity-size relation is found to be strongly dependent on galaxy environment.
This variation is in the sense that, at a given morphology, larger  and more luminous galaxies are rarer in sparser environments. 
Our analysis of late-type galaxy morphologies reveals that scatter increases towards later Hubble types.
Taken together, these results place strong constraints on conventional hierarchical models in which galaxies are built up in an
essentially stochastic way. 

\end{abstract}
 

\section{INTRODUCTION}

\label{sec:introduction}

In hierarchical models the formation of galaxies is driven by multiple mergers and complicated feed-back effects. As a result the star-formation and mass-building history of individual galaxies can differ in a multitude of ways. However, a common feature of these models is that the end-product of this process is generally a system which lies on (or at least near) a fundamental scaling relation e.g. the fundamental plane for elliptical galaxies \citep{BoylanKolchin:2006p21539,Robertson:2006p15098,Ciotti:2007p22639,Covington:2008p1439}, or the Tully-Fisher relation for spiral galaxies \citep{Dalcanton:1997p21815}). In some models this agreement is somewhat artificial, because the form of the scaling relation is taken as an input parameter, but in other cases it emerges because some fundamental relations are closely linked to simple underlying physics (e.g. the Faber-Jackson relation emerges from the virial theorem coupled with assumptions about mass-to-light ratio being a weak function of mass).

How can measurements of the fundamental scaling relations of galaxies best be used to constrain ideas for galaxy formation? It is arguable that the most direct way forward is simply to measure how the form of these relations changes with environment.  Environment is the central parameter in hierarchical models, since in this picture the ultimate fate of a galaxy is determined mainly by its merger history, and the rate of merging is accelerated in richer environments, which are expected to host the oldest and most massive galaxies at any time \citep{Mo:1996p22126}. It is perhaps even more interesting to measure the {\em scatter} about the fundamental relations as a function of environment, because even if a relation's mathematical form is set by simple physics, the process by which a galaxy winds up on that relation is governed by random processes. A galaxy's stochastic merger history sets the timing over which it builds, but its final state is the product of a complex interplay between a large number of parameters, such as its initial mass, angular momentum and dark halo concentration, not to mention feedback to the intergalactic medium, so it would be surprising if the scatter about the fundamental relations were very small.

In a recent paper \cite{vandenBergh:2008p11891} found that the luminosity-diameter relation provides the tightest of all the purely photometric correlations used to characterize galaxies. This fact was first noted by \cite{Giuricin:1988p12909}, who also claim to have detected ``appreciable differences in the galaxy-luminosity relationships for different clusters''. On the other hand,  \cite{Girardi:1991p12914} found no significant differences between the luminosity-diameter relations in a variety of environments. \cite{Gavazzi:1996p21749} and \cite{vandenBergh:2008p11891} have reached similar conclusions, albeit with small samples, and both suggest that galaxies are best viewed as complex systems linked by a single fundamental property, namely stellar mass. This viewpoint is supported by  \cite{Disney:2008p12920}, who concluded the same thing on the basis of a principal component analysis of data from a sample of galaxies selected from a large blind survey of neutral hydrogen gas emission.

On the other hand, recent work \citep{Shen:2003p70,Kauffmann:2003p7199,Bernardi:2003p1624,Desroches:2007p19582,Hyde:2008p12893} using much larger samples drawn from the SDSS \citep{York:2000p3192} have shown the size-luminosity and size-mass relations exhibit curvature for both early and late type galaxies. In addition, \cite{Bernardi:2007p1618,Bernardi:2009p21529,Lauer:2007p21894,vonderLinden:2007p21830,Liu:2008p22026,Coenda:2009p18836} have found that the central or brightest cluster galaxies (BCGs) exhibit a steeper size-luminosity relation than field early type galaxies, though \cite{Guo:2009p19543} and \cite{Weinmann:2009p19068} do not find the trend. 

The works just described are based on either small ($<$1000) samples of visually classified objects or much larger samples ($>$50,000) with very broad morphological segregations based on measurements of central concentration. The recent publication of the detailed morphological catalog of \cite{Nair:2010p22790} offers us an opportunity to revisit the question of the environmental dependence of the galactic scaling relations as a function of morphology and environment using samples with high-precision classifications that are an order of magnitude larger than those available to previous authors. Our aim in the present paper is to determine whether the form and scatter of the luminosity-size relation for galaxies remains constant as a function of environment. We will explore a range of Hubble types from E to Sc, but our main emphasis will be on early-type systems. This is because in hierarchical models elliptical galaxies are the end-product of major mergers, so they are a crucible for testing the central idea of merger-driven evolution. Early-type galaxies also provide us with the closest connection with existing work on this subject, because they dominate the populations of rich clusters, and most previous work on the environmental dependence of the luminosity-size relationship has focused on simple comparisons between rich clusters and the field. Our own sample differs from this earlier work in that it does not focus on rich clusters, and instead spans a range of environments from sparse regions in the field up to systems that could best be characterized as poor-intermediate clusters. In our study there are only 700 galaxies in clusters with $\log(M_{halo}/ M_\odot)>14.0$ and our data includes galaxies
from only 47 clusters with more than 50 members (described below). 

The plan for this paper follows: Our sample is described in Section~\ref{sec:sample}, followed by our measurements of the fiducial luminosity-size relationships in Section~\ref{sec:LR}. 
The dependence of these relationships on environment is explored in Section~\ref{sec:LRenv}.
Section~\ref{sec:LRcc} looks into the importance of color and central concentration on scaling relations.
Section~\ref{sec:BCG} investigates the luminosity-size relationship for the brightest cluster galaxies (BCGs) in comparison to field and cluster satellite galaxies. Our results are discussed in Section~\ref{sec:discussion}, and conclusions are presented in Section~\ref{sec:conclusion}. We compare the two SDSS size measures, R(90) and R(dev), in Appendix A and compare the publicly available environment estimates used in this paper in the Appendix B.
Throughout this paper we assume a flat dark energy-dominated cosmology with $h=0.7$ and $\Omega_\Lambda$=0.7. All magnitudes quoted are in the AB system.

\section{Sample}
\label{sec:sample}

We use the sample of  $14034$ visually classified bright galaxies from \cite{Nair:2010p22790} (NA10) to investigate the luminosity-size relation in the local universe. The reader is referred to the NA10 catalog paper for details, but in summary the local sample is derived from the SDSS  DR4 release which covers 6670 square degrees in the \em{u',g',r',i',z'} \rm  (3553 \AA , 4686 \AA, 6166 \AA, 7480 \AA, 8932 \AA) bands. The galaxies are selected from the spectroscopic main galaxy sample described in \cite{Strauss:2002p6172}. The DR4 best photometry catalogs were used to select all objects with an extinction corrected {\em g'}-band magnitude brighter than 16.0 at redshifts between 0.01 and  0.1. Visual classification of the entire sample
was carried out by one of the authors (PN) using the Carnegie Atlas of Galaxies \citep{Sandage:1994p4888} as a visual training set, in consultation with the Third Reference Catalog of Bright Galaxies \citep[RC3]{deVaucouleurs:1991p4597}, along with images for many fiducial objects obtained using the IPAC NED database. The classifications are found to agree with those from the RC3, with a mean deviation of 1.2 T-types for the $\thicksim$ 1700 galaxies in common to both samples. It is important to note that although our sample probes a wide range in mass and luminosities, it has magnitude cuts imposed and is not volume-limited. However, this sample can be used to understand the influence of detailed morphology as well as to compare the different metrics used to measure morphology such as the S\'ersic index, central concentration and color. For this paper, we use the SDSS derived g-band Petrosian R(90) radii and absolute magnitude \citep{Strauss:2002p6172}, corrected for galactic extinction \citep{Schlegel:1998p10925} and K-corrected to z=0.0 using {\em K-correct} \citep{Blanton:2005p79}. The effect of seeing correction on the Petrosian measure of sizes is negligible (see Appendix A). It should be noted that Petrosian magnitudes are not ideal because they capture a type-dependent fraction of the total light of a galaxy \citep{Graham:2005p19601,Bernardi:2009p22173}. However, unlike de Vaucouleur sizes, they are not sensitive to central concentrations within a type (see Appendix A). We use stellar masses derived by \cite{Kauffmann:2003p97}. The sample analyzed here is a clean sub-sample of $12150$ objects with no projection effects by satellites, foreground objects or nearby stars. This leads to the exclusion of $\sim 400$ BCG and some nearby satellite galaxies as defined by \cite{Yang:2007p19054}. Environment estimates are available for the entire sample. The environmental estimates used will be described next.

\begin{figure*}[t!]
\rotatebox{0}{\resizebox{18cm}{5.5cm}{\includegraphics{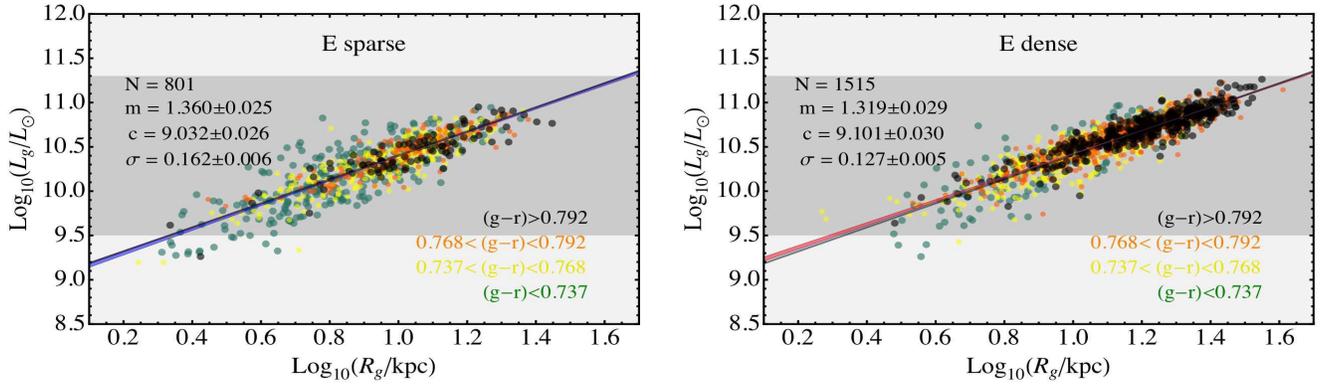}}}
\caption[Plot]{\label{fig:LvsRforE}The g-band luminosity versus size plot, for galaxies in the sparsest regions
(left-hand panel) and densest regions (right-hand panel) in our sample. Luminosity is in solar units,
and size is defined using the Petrosian R(90) radius in the rest frame.
Sparse regions are defined as N$\le$3 and Baldry rho $<$ -0.32, while dense regions are defined
as N$\ge$3 and Baldry rho $>$ -0.32.
The points are keyed to $g-r$ color quartiles. Black points are the reddest galaxies in each panel, followed by orange, then yellow with green points being the bluest galaxies. 
Note the remarkably low scatter about the luminosity-size relationship.
See text for details.
}
\end{figure*}

\subsection{Measures of Environment}

There are multiple metrics that have been used to describe the environment of galaxies in the Sloan Survey, such as
the over-density estimates used by \cite{Blanton:2005p2271}, an Nth nearest neighbor approach 
used by \cite{Baldry:2006p103}, and group catalog algorithms used by \cite{Yang:2007p19054}. 
What follows is a short summary of these three main techniques, and some justification for
our decision to only use two of these (the Nth nearest neighbour aproach of \cite{Baldry:2006p103}, and the
group catalog approach of \cite{Yang:2007p19054} in the present paper. 
The reader is referred to the original papers cited above for details.

The  environmental overdensity $\rho$ estimated by the New York Value Added Galaxy Catalog (NYU-VAGC) group \citep{Blanton:2005p2271} was calculated for each spectroscopically targeted galaxy in the SDSS (14.5$<$r$<$17.77 and 0.05$<$z$<$0.22) 
For each galaxy, neighbors were counted in the `SDSS imaging in the magnitude range corresponding to M$^{*}\pm$1 mag (passively evolved and K-corrected as for an early-type galaxy) and within 5 h$^{-1}$ Mpc (transverse, proper)', excluding the target galaxy. The count was weighted  to recover the estimated overdensity averaged over a spherical three-dimensional Gaussian window $e^{-r^{2}}/2a^{2}$ with a radius of a = 1 h$^{-1}$ Mpc (proper).

\cite{Baldry:2006p103} measured the environmental density for  SDSS galaxies with an extinction corrected r-band magnitude brighter than 18  for spectroscopically selected galaxies between  0.005$<$z$<$0.3 and photometrically selected galaxies with surface brightness 18.5$<$$\mu_{r,50}$$<$24.0. The density is defined as  $\Sigma = N/ (\pi d_{N}^{2})$,
where $d_{N}$ is the projected comoving distance (in Mpc) to the Nth nearest neighbour (within $\pm$1000 km/s if a spectroscopic redshift was available or else with photometric redshift errors within the 95 percent confidence limit). 
A best estimate density (to account for spectroscopic incompleteness) was obtained by calculating the average density for N=4 and N=5 with spectroscopically confirmed members only and with the entire sample. The mean Log $\Sigma$ for our sample is -0.32.

\cite{Yang:2007p19054}  used an iterative halo-based group finder on the NYU-VAGC SDSS catalog for objects with an extinction corrected r-band magnitude brighter than 18.0 mag and 0.01$<$z$<$0.2 with a redshift confidence $C_{z}$$>$0.7. 
Tentative group members were identified using a modified friends-of-friends algorithm. The group members were used to determine the group center, size, mass and velocity dispersion. New group memberships were determined iteratively based on the halo properties. The final catalog yields additional information identifying the brightest galaxy in the group (BCG), the most massive galaxy in the group (both used as proxies for central galaxies), estimated group mass, group luminosity and halo mass. \footnote{For groups with only one member, the brightest galaxy and most massive galaxy flags are also set.}

As described in the Appendix B, when comparing the different environmental estimates we find the \cite{Blanton:2005p2271} measurements are biased against small galaxies in dense environments in comparison to the other two estimate. Hence, in this paper we use the \cite{Baldry:2006p103} and  \cite{Yang:2007p19054} estimates to define galaxies in clusters and in the field. We use the group occupation number N as a proxy for environment. Group mass, luminosity or distance from the group center can also be used. 
In the Appendix we show a comparison between the \cite{Baldry:2006p103} environmental estimator and the group occupation number determined by  \cite{Yang:2007p19054}. We find that for groups with more than 7 spectroscopically confirmed members, the Yang and Baldry estimates of dense environments are consistent. However, galaxies considered to be more  isolated by Yang (N$<$3) can occur in a range of environments as determined by \cite{Baldry:2006p103}. This emphasizes the difficulties in determining environment (small scales vs group scales) and which measure matters \citep{Blanton:2007p19165,Deng:2009p20426}. We are primarily interested in relative evolution between field and cluster galaxies and hence define low density regions as galaxies with N$\le$2 and Baldry rho $<$ -0.32 while high density regions are defined as galaxies in groups with more than two members N$>$2 and Baldry rho $>$ -0.32 (the mean N and Log $\Sigma$ for our sample). When we consider BCGs versus satellite galaxies in clusters, we do not include groups with occupation numbers less than 2 unlike \cite{Weinmann:2009p19068} who assume satellite galaxies were not detected for those `central' galaxies with N$<$2.

\begin{figure*}[htbp!]
\unitlength1cm
\hspace{0.5cm}
\rotatebox{0}{\resizebox{17cm}{15cm}{\includegraphics{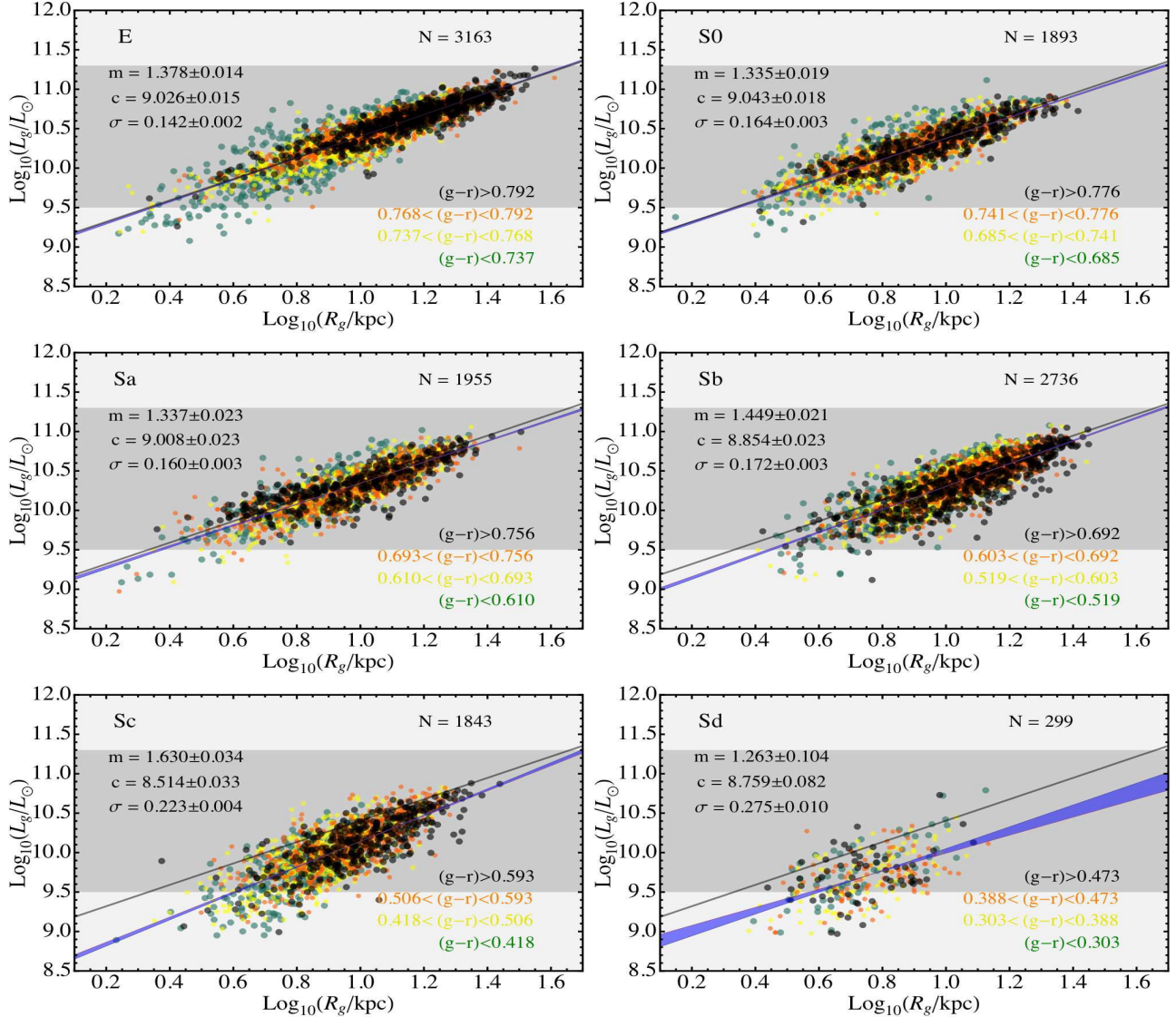}}}
\caption[Plot]{\label{fig:LvsRperTType} The g-band luminosity versus size for a range of galaxy types. Luminosity is in solar units, and size is defined using Petrosian R(90) radius in the rest frame. Each panel isolates a specific Hubble type \cite{Nair:2010p22790}, as noted in the text. The points are keyed to $g-r$ color quartiles. Black points are the reddest galaxies in each panel, followed by orange, then yellow with green points being the bluest galaxies. The black line shows the relation for elliptical galaxies.  The blue lines/regions in each panel denote the best fit relation for each particular Hubble type.
}
\end{figure*}

\section{Luminosity-Size Relations}
\label{sec:LR}

The luminosity-size relationship for early type galaxies is shown Figure~\ref{fig:LvsRforE} for sparse and dense environments.
The points are keyed to color quartiles for each Hubble type where black points are the reddest galaxies in each panel and green points the bluest galaxies. Orange and yellow points are the intermediate quartiles. The shaded region in each plot shows the luminosity range spanned by elliptical galaxies in dense regions. 
This figure clearly shows that the luminosity-size relationship of elliptical galaxies defines
a remarkable tight relationship. 
For elliptical galaxies as a whole, the data are well-represented by a power-law 
of the form:

\begin{equation}
\log(L_{g}) = (9.03\pm0.02 ) + (1.38\pm0.01)\times\log(R_{p90}).
\end{equation}    
\vspace{0.1cm}
    
\noindent This equation defines the fiducial line shown in black on both panels of 
Figure~\ref{fig:LvsRforE}. 
The colored sector in each panel of this figure
illustrates the $\pm 1\sigma$ range of permissible slopes. 
For high-density environments (rho$>$-0.32 and N$>$2)  the data for E+E/S0 galaxies are
well-represented by the equation:\\
\begin{equation}
\log(L_{g}) = (9.10\pm0.03 ) + (1.32\pm0.03)\times\log(R_{p90}).
\end{equation}    
\vspace{0.1cm}

\noindent The corresponding relation for E+E/S0 galaxies in low-density environments (rho$<$-0.32 and N$\le$2)
is given by:\\
\begin{equation}
\log(L_{g}) = (9.03\pm0.03 ) + (1.36\pm0.03)\times\log(R_{p90}).
\end{equation}    

We therefore find that {\em  the luminosity-radius relations for E galaxies
are statistically indistinguishable for high and low density regions}.
However, inspection of Figure~\ref{fig:LvsRforE} and of Table~\ref{tab:relations} shows that the
the RMS dispersion about these relations do show an environmental
dependence\footnote{Note that dispersions in Table~\ref{tab:relations} are in dex, not magnitudes.}.
The dispersion around the fiducial relation is $\sigma=0.13$ dex
in dense environments, and $\sigma=0.16$ dex in sparse environments
(equivalent to 0.32 mag and 0.40 mag, respectively). In agreement with work by
\cite{Gallazzi:2006p2}, who investigated the environmental dependence
of the Faber-Jackson relation with SDSS galaxies (though with a larger dispersion of $\sigma=0.58$ mag), 
we find the reddest elliptical galaxies lie
predominantly in dense regions, and that these show a smaller dispersion 
than do corresponding systems that are blue (and which lie in
sparser regions). What is surprising, however, is that the dispersion
about the fiducial relationship is so small. 

For the elliptical galaxy sample as a whole (including galaxies in all environments) 
shown in Figure~\ref{fig:LvsRperTType}, 
the dispersion is only 0.14 dex (0.35 mag). The significance of this very
small dispersion about the luminosity-size relationship is best appreciated 
by comparing it with the corresponding scatter about the fundamental plane.
\cite{Bernardi:2006p1613} have derived the dispersion in the distance from the 
fundamental plane in $g$-band for early-type galaxies, and their Table~5 
shows a dispersion of $\sigma=0.345$ mag  in high-density regions 
and $\sigma = 0.355$ mag in low-density regions, defined
over a sample spanning a similar dynamic range of luminosity
to that being examined here. 

Therefore {\em the tightness of the
luminosity-size relationship for elliptical galaxies
is comparable to the scatter about the fundamental plane}. 
This is in spite of the fact that
the luminosity-size relationship is defined on the basis
of {\em purely photometric parameters}, with no spectroscopic
information needed (aside from a redshift). The luminosity-size
relation exhibits a scatter ($0.35$ mag)
that is almost half the typical scatter
in the Faber-Jackson relation ($\sim0.58$ mag, \cite{Gallazzi:2006p2}).

The remarkable tightness of the luminosity-size relationship should not
blind us to the existence of some significant scatter about the
fiducial line given by Equation (1), and
it is interesting to consider the source of this dispersion. 
We will have more to say about this in Section~5,
and for now only note that it is apparent from 
Figure~\ref{fig:LvsRforE} that the reddest elliptical galaxies (black points)
exhibit the tightest relation between size and luminosity
(in both dense and sparse environments). This suggests that
star-formation history (contributing through flux-weighted stellar population
age) is an important component of the scatter, as has been found by \cite{Shankar:2009p22206}. We
will show later that the dispersion in the relationship can be
decreased even further by incorporating central concentration as a third
parameter, suggesting that both star formation history and structure are important
contributors to the scatter.

Expanding our consideration to later Hubble types, Figure~\ref{fig:LvsRperTType} shows the 
distributions of galaxy size vs luminosity in panels which segregate galaxies into the following 
broad type bins: E+E/S0, S0+S0/a, Sa+Sab, Sb+Sbc, Sc+Scd, Sd+Sm. 
Linear least-squares fits to the relations shown in this figure are included in  Table~\ref{tab:relations}. 
In the remainder of this paper we will drop the sub-type nomenclature when referring
to galaxies, and thus when we refer to elliptical galaxies we refer to E + E/S0 galaxies.
It is seen from Figure~\ref{fig:LvsRperTType} and Table~\ref{tab:relations} that both the slope and the
dispersion in the best-fitting relationships increases with morphological type (i.e with later morphologies), 
with fairly little dependence on the color of the galaxy. 
To allow the various panels to be compared more easily, the best fit relation for elliptical galaxies is shown in every panel of
Figure~\ref{fig:LvsRperTType} as a black line. The blue lines/regions in each panel denote the best fit relation for each particular Hubble type. Our aim is to allow the reader to determine from inspection of this figure whether the slope of the best-fitting line differs from that defined by the elliptical galaxy population. To this end, for each Hubble type we have plotted
the lines with maximum and minimum slope consistent with the $1\sigma$ uncertainties (determined by 100 bootstraps), and
shaded the region between these lines in light blue.  

It is apparent that as Hubble type increases an increasing fraction of objects
fall below the relationship for elliptical galaxies defined by Eqn. (1).  As has already been noted,
the values of the slope and intercept of the best linear fits to the $L_g$ vs. $R_{p90}$ 
data for each Hubble type are presented in  Table~\ref{tab:relations}.
The uncertainties on the tabulated quantities presented in this table were all
calculated using the statistical bootstrap method (Efron \& Tibshirani 1994)
with 100 iterations. This table also contains fit data for sub-categories of galaxies 
grouped by environment, as described in the next section. 

\section{Effect of Environment}
\label{sec:LRenv}

In our discussion of Figure~\ref{fig:LvsRforE} it was already clear that environment did not
change the form of the fiducial luminosity-size relationship for early-type
galaxies, but that it did play a significant (but not overwhelming)
role in increasing (slightly) the dispersion about the fiducial relationship.
It is also clear that environment plays a role
in skewing the early-type population as a whole toward brighter galaxies as
the density increases. In this
section we examine whether similar trends are seen in the sample as a whole. Before
doing this, it is first important to note that
the dependence of galactic properties on the environment of galaxies has recently been studied 
by various groups using different metrics to define  environment and morphology. 
\cite{Blanton:2003p2277}, \cite{Blanton:2005p2271} and \cite{Hogg:2004p294} find that luminous galaxies
reside preferentially in high-density regions and that blue galaxies
are mainly located in low-density regions. 
\cite{Coenda:2009p18836}
find that bright galaxies preferentially inhabit cluster centers, and
that early-type galaxies in the field have lower luminosities than do their
counterparts in clusters. 

\begin{figure*}[htbp!]
\unitlength1cm
\hspace{0.5cm}
\rotatebox{0}{\resizebox{16cm}{23cm}{\includegraphics{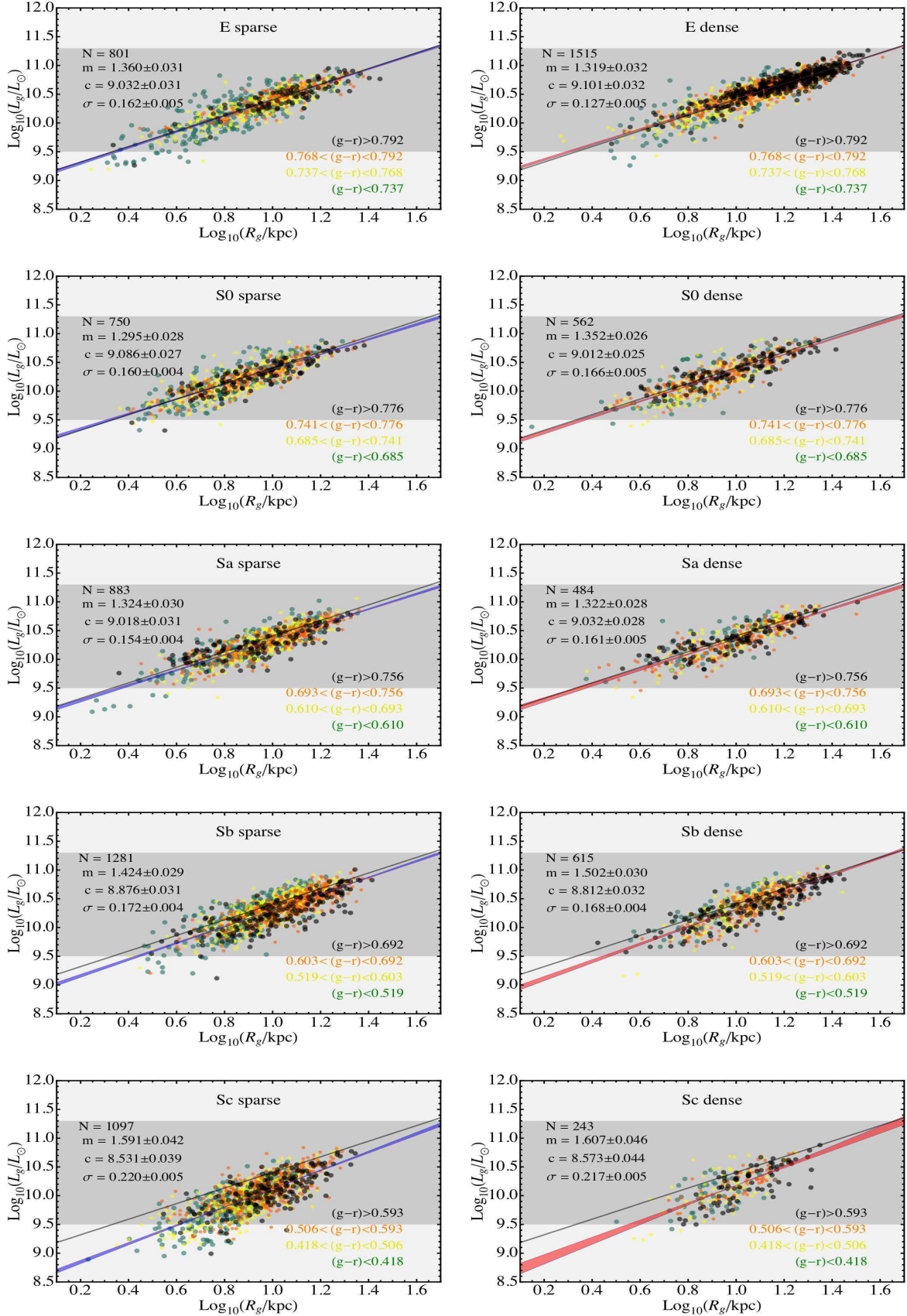}}}
\caption[Plot]{\label{fig:LvsRperTTyperho} g-band luminosity versus g-band Petrosian R(90) radius in kpc per morphological type in low density regions (left) and high density regions (right). The points are keyed to $g-r$ color quartiles. Black points are the reddest galaxies in each panel, followed by orange, then yellow with green points being the bluest galaxies. The black line shows the relation for elliptical galaxies.  The blue lines/regions and red lines/regions in each panel denote the best fit relation for each particular Hubble type in sparse and dense environments respectively.
}
\end{figure*}

In Figure~\ref{fig:LvsRperTTyperho} we now investigate these effects in greater detail
using the environmental density parameters described earlier.
(Note that these density estimates are all contained in the NA10 catalog).
For simplicity, we once again adopt the strategy used in Figure~\ref{fig:LvsRforE} and divide the
data for each morphological type into two bins corresponding to regions
of above-average, or below-average, environmental density.

\begin{figure*}[htbp!]
\unitlength1cm
\hspace{0.5cm}
\rotatebox{0}{\resizebox{16cm}{23cm}{\includegraphics{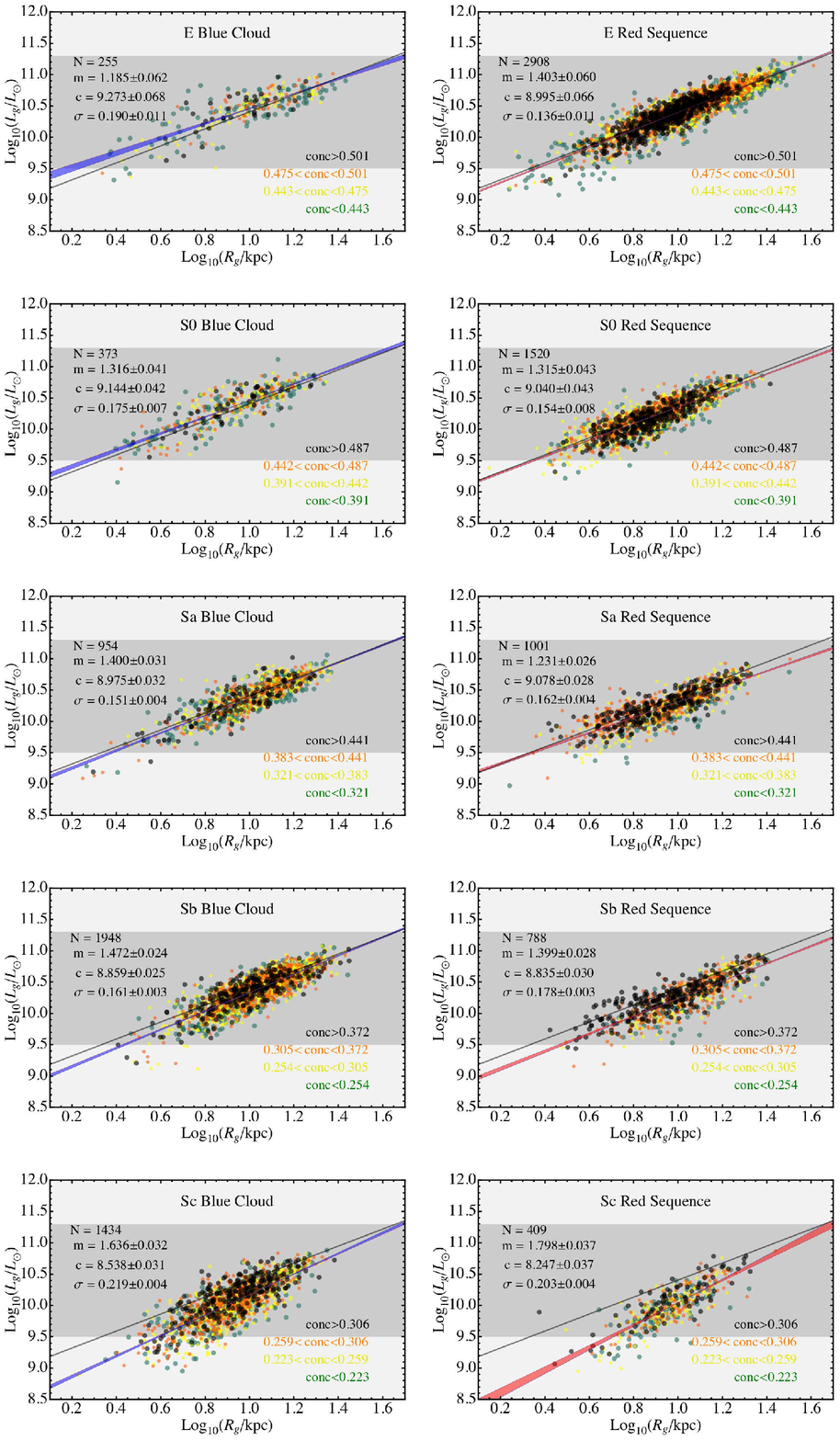}}}
\caption[Plot]{\label{fig:LvsRperTTypeCM} g-band luminosity versus g-band Petrosian R(90) radius in kpc as a function of morphological type in the blue cloud (left) and red sequence (right) keyed to central concentration. The most concentrated galaxies are shown in black, and the least concentrated are shown in green. Orange and Yellow are intermediate concentrations with orange points sampling the larger quartile. The black line shows the relation for elliptical galaxies.  The blue lines/regions and red lines/regions in each panel denote the best fit relation for each particular Hubble type in the blue cloud and red sequence respectively.
}
\end{figure*}

Figure~\ref{fig:LvsRperTTyperho} illustrates how the trends shown in the
previous figure depend on environment, and the main message seems to be that 
environment has little effect on the shape of the fiducial relationships for each type, 
although, as with elliptical galaxies, there appears to be a somewhat greater scatter 
about the relationship in spare environments. 
For example, the dispersion of the data around the fiducial relationship is found to be
$\sigma=0.22$ dex for Sc galaxies. This value is much larger than that for E
($\sigma = 0.14$ dex) and Sa ($\sigma = 0.16$ dex) galaxies.  
The relationship between scatter and galaxy color noted in our discussion of 
Figure~\ref{fig:LvsRforE} turns out to be generally true, as we see that the
scatter is smallest amongst red galaxies for every Hubble type.

An arguably more important effect of environment is also shown
in Figure~\ref{fig:LvsRperTTyperho}. Environment seems
to curtail the range of sizes exhibited by certain Hubble types,
while leaving the form of the luminosity-size relation unchanged.
This is made clear by inspection of the final two
columns in  Table~\ref{tab:relations}, which lists the size range
spanned by galaxies from the 10\%--90\% quantiles.
The absence of large ellipticals in the field has already been
noted, but another effect of environment is to limit the number of
small Sc galaxies in rich environments. This effect will be discussed further in the 
next section.

\begin{figure*}[t!]
\unitlength1cm
\hspace{0.5cm}
\begin{minipage}[t]{4.0cm}
\rotatebox{0}{\resizebox{16cm}{5cm}{\includegraphics{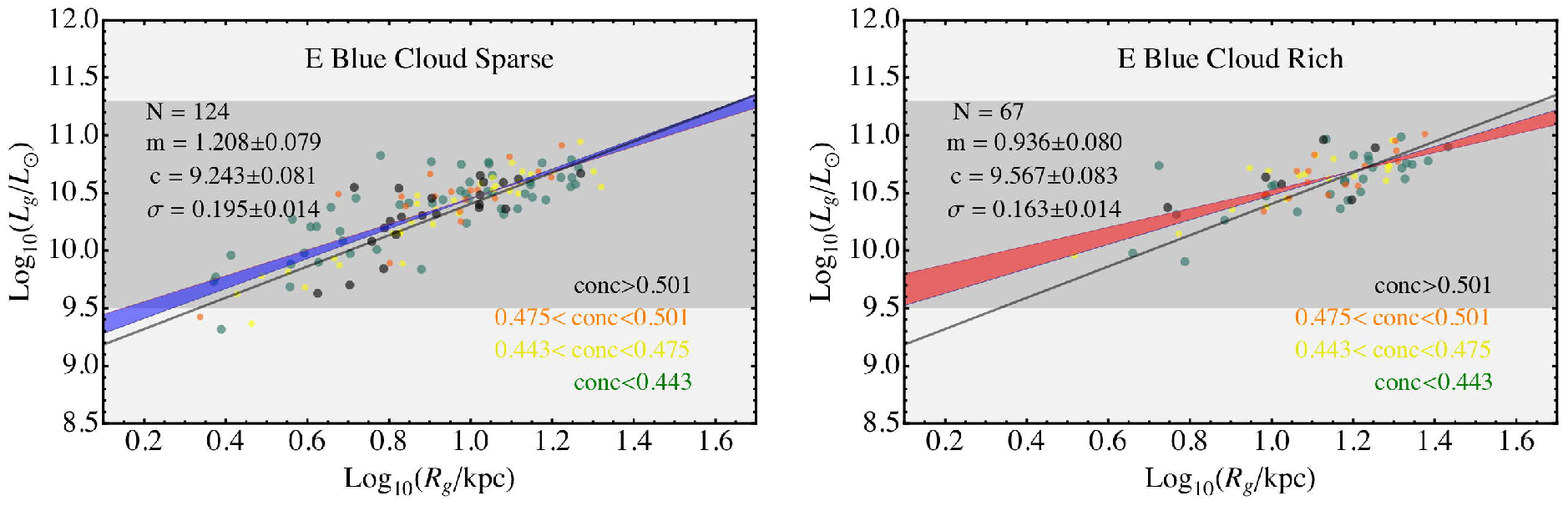}}}
\rotatebox{0}{\resizebox{16cm}{5cm}{\includegraphics{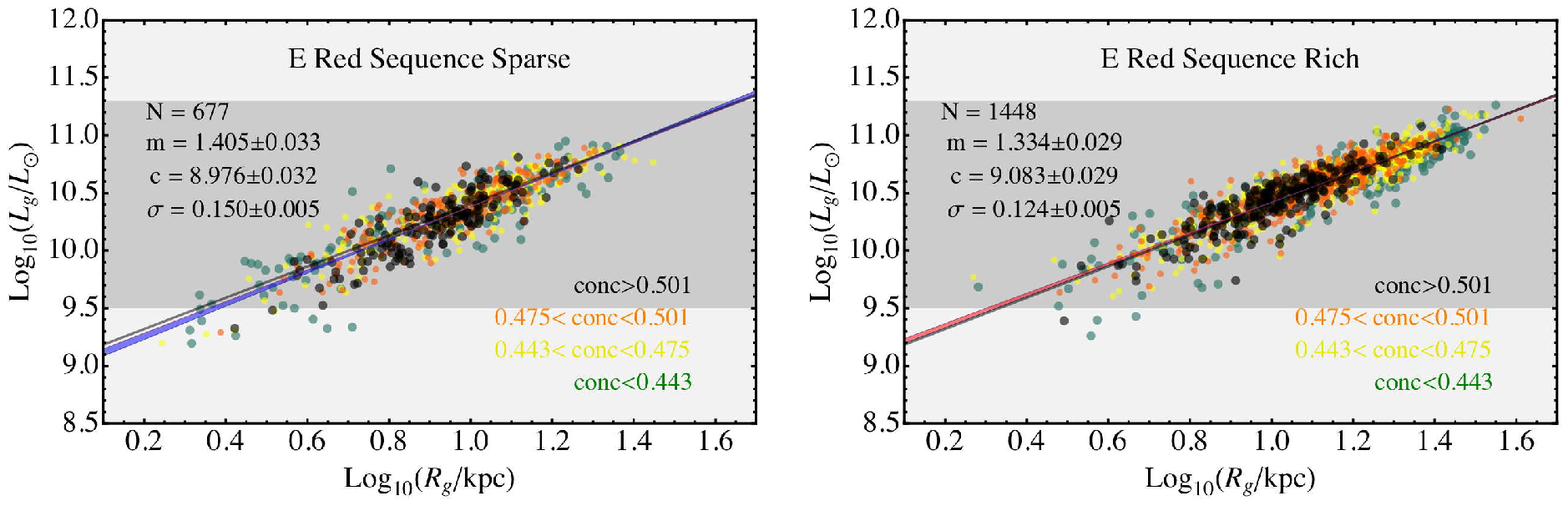}}}
\end{minipage}
\caption[Plot]{\label{fig:LvsREredblueRHO}  g-band luminosity versus g-band Petrosian R(90) radius in kpc per morphological type on the blue cloud (top) and  red sequence (bottom) in low (left) and high density (right) regions keyed to central concentration. The most concentrated galaxies are shown in black, followed by orange and yellow with the least concentrated galaxies shown in green.
  }
\end{figure*}

\section{Effect of color and concentration}
\label{sec:LRcc}

Figure~\ref{fig:LvsRperTTypeCM}  shows galaxies defined to be on the red sequence or the blue cloud (as defined by
\cite{Baldry:2004p43}) for each morphological bin keyed to the central concentration\footnote{Defined as the ratio of flux within an inner and outer elliptical aperture determined from the sky-subtracted, intensity-weighted, second-order moment of the image. 
 The major and minor axes of the outer aperture are normalized so that the total area within the ellipse is the area of the galaxy. The inner aperture is defined by scaling these axes down by a factor of 3.} of the galaxy. 
A number of striking trends are obvious from this figure, and from the summary of the trends given in Table~\ref{tab:relations}.
Firstly, it is seen that the scatter in the luminosity-size relation for elliptical galaxies is somewhat
smaller on the red sequence than in the blue cloud. From Figure~\ref{fig:LvsRforE} we find that color
does matter in terms of increasing the dispersion, and Table~\ref{tab:relations} indicates that the scatter
is $\sim50$\% larger for early-type galaxies in the blue cloud than on the red sequence.
Secondly, we find increasing central concentration also
has the effect of decreasing the dispersion and making modest changes
in the slopes and intercepts of the fiducial relation for elliptical galaxies.
This trend is explored in greater detail in Table~2, which investigates the luminosity-size
relationships for early-type galaxies as a function of concentration. 
We see from this table that the slope and intercept of the luminosity-size relation for
high-concentration early-type galaxies appears to be significantly different than for
low-concentration early-types. This is manifested as a mild stratification in the plot
symbol colors in the top-right panel of Figure~\ref{fig:LvsRperTTypeCM}.
Our suspicion is that the trends with concentration  may well be due to merger history, following on ideas given by
\cite{Kormendy:2009p21817}, who find that the merger history of elliptical galaxies 
determines the presence of internal structures like cores and excess light in the central 
regions of galaxies as well as their S\'ersic index\footnote{\cite{Kormendy:2009p21817} also
suggest that ``extra light ellipticals got their low S\'ersic indices by forming in relatively few binary mergers, whereas giant ellipticals have $n > 4$ because they formed in larger numbers of mergers of more galaxies at once plus later heating during hierarchical clustering.''  Interestingly, these authors also find that core Es contain X-ray-emitting gas whereas extra light Es generally do not.}. 
On the other hand, since the general trends remain similar and the trends
with concentration seem a second-order effect, 
our results (and the work of \citep{vandenBergh:2008p11891})
seem to suggest that the overall
dimensions of early type galaxies ({\em i.e.} sizes measured at  a faint isophote, well
beyond the inner regions)
do not seem to be greatly affected by mergers.
The luminosity-size relation for Sa - Sc galaxies does show a 
mild dependence  when moving from the blue cloud to the red sequence, 
such that at a given luminosity, quenched, red sequence galaxies are slightly larger.
In addition, it is interesting to note that late type galaxies in Figure~\ref{fig:LvsRperTTypeCM}
do not show as strong a trend  with concentration as elliptical galaxies, 
possibly indicative of fewer mergers in their formation history, or perhaps of
more minor mergers. 

The environmental dependence of the trends just noted are explored in
Figure~\ref{fig:LvsREredblueRHO}, which  shows the luminosity versus size relation for blue cloud (top) and red sequence (bottom) elliptical galaxies in low (left) and high density (right) regions keyed to central concentration. Blue cloud elliptical galaxies are found in both low and high density regions, though there are very few small, low luminosity ellipticals in rich environments.  We find that blue cloud elliptical galaxies exhibit significantly different slopes than do red sequence galaxies. In addition, red sequence elliptical galaxies exhibit very little change in moving from low to high density regimes, 
unlike blue cloud galaxies which exhibit a shallower luminosity versus size relation than the general elliptical population. 
Given the small numbers of blue elliptical galaxies present in our sample ($<$200), they do not influence the overall luminosity size relation of elliptical galaxies. However at higher redshifts, the blue elliptical population may be significant in both low and high density regimes and may cause a shift in the overall luminosity size relation. Any high redshift study which identifies ellipticals by using the red sequence technique or color cuts \citep{Trujillo:2004p17343,Trujillo:2006p151} is likely to miss this population of galaxies.

In summary, Figure~\ref{fig:LvsRperTTypeCM} 
and Figure~\ref{fig:LvsREredblueRHO} 
appears to show that variations in concentration (i.e. homology) 
as well as variations in color (i.e. age) combine to define
the scatter about the luminosity-size relationship. The
visual impression from Figures~4 and 5, backed up by the
results summarized in Tables~1 and 2, is that
variations in
homology are manifested in small changes the slopes and intercepts of the
fiducial luminosity-size relation for early-type galaxies.
On the other hand, comparing the variation as a function of color in the {\em range} spanned by
galaxies at various Hubble types suggests that color (i.e. age) plays the defining role
here. For example, it is seen that for early-type galaxies the reddest systems
tend to be largest and most luminous, and that a related trend appears to hold for
late type galaxies, few of which are both red and under-luminous.
Taken together, these observations seem to provide some 
evidence for the notion that the origin of the scatter in the luminosity-size relation can be
described in the following simple terms: for a given Hubble type,
variations in homology change the 
slope and intercept in the local luminosity-size relationship,
while variations in age truncate the size range spanned by galaxies.

\section{Is the cluster center important?}
\label{sec:BCG}

In the previous sections we considered the dependence of  the luminosity-size relation on global environment  and color. Although global environment does influence the mass function of individual types, it is not the only useful measure for testing hierarchical formation models. In fact, the position of a galaxy within a cluster may be more important than global density,
since different physical processes occur in the outskirts versus the centers of groups or clusters \citep{Dressler:1980p10614}.  \cite{BoylanKolchin:2006p21539} find that the merger history of BCGs,  where mergers preferentially occur on radial orbits, would lead to a steeper size-luminosity relation compared to field galaxies. In addition, satellite galaxies could lose their extended gas haloes (starvation/strangulation, e.g. \cite{Larson:1980p21930}) by interaction with the intracluster medium and ram pressure stripping \citep{Gunn:1972p17919}. These two processes are often cited as being important to the creation of S0 galaxies in clusters, though it is hardly an exhaustive list, and a number of other processes could be responsible for the
requisite gas loss (e.g. there is evidence that an AGN might be responsible for turning NGC3115 into an S0, c.f. \cite{vandenBergh:2009p21551}).

Recent studies with the SDSS have found conflicting results with respect to size-luminosity relationship of BCGs. \cite{Lauer:2007p21894}, \cite{Bernardi:2007p1618}, \cite{Liu:2008p22026}, \cite{Coenda:2009p18836} and \cite{Bernardi:2009p21529},
using the C4 cluster catalog \citep{Miller:2005p22112} and/or the MaxBCG cluster catalog \citep{Koester:2007p22118}, 
find that early-type BCG galaxies are larger at fixed luminosities and show a steeper size-luminosity (Y versus X) relation than satellite or field galaxies. On the other hand, \cite{Guo:2009p19543} and \cite{Weinmann:2009p19068} using the \cite{Yang:2007p19054} halo catalog find no difference between central and satellite galaxies in the size-luminosity plane. 

\begin{figure}[t!]
\unitlength1cm
\rotatebox{0}{\resizebox{8.5cm}{6cm}{\includegraphics{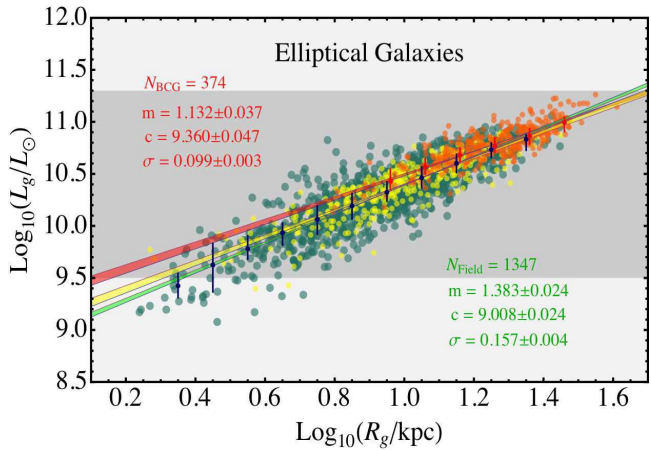}}}
\caption[Plot]{\label{fig:LvsRBCG} g-band luminosity versus g-band Petrosian R(90) radius in kpc for E and E/S0 galaxies showing the brightest cluster galaxies ( BCG; orange points), satellite galaxies (yellow points) and field galaxies (green points). Best fit lines for each environments are shown in red, yellow and green. The red (green) lines show the range spanned by the first and third quantile in each luminosity bin. 
}
\end{figure}

Figure~\ref{fig:LvsRBCG} shows the luminosity-size relation for elliptical galaxies in our sample for field galaxies (in green), satellite galaxies in groups with 10 or more members (in yellow) and BCG galaxies in groups with ten or more members (in orange). Best fit lines for each sample are over-plotted. Points with error bars indicate the 25th and 75th quantiles in various
size bins. We find that BCG galaxies have a shallower size vs luminosity slope (steeper luminosity vs size slope) consistent with \cite{Bernardi:2009p21529} and others. 
This result may be due to some mild curvature in the luminosity-size relationship, since the BCGs span such a limited range of luminosities (which is why, of course, they have been used as distance indicators), and in any case the slope is difficult
to define with confidence because of the small luminosity range spanned.  
However, it is interesting to note that quantile distribution of luminosities (blue and red error bars) is similar for field and BCG galaxies, where the samples overlap, especially at higher luminosities, implying little difference between BCGs and field galaxy scaling relations in the region of
overlap.  BCGs show a tighter luminosity size relation than the general elliptical galaxy population with a dispersion of only $0.25$ mag. Satellite cluster galaxies show a slightly larger dispersion while field galaxies show the largest dispersion (0.39 mag). This is consistent with previous studies by \cite{Bernardi:2007p1618} and \cite{Bernardi:2009p21529}. These authors suggest that the small dispersion for BCG galaxies indicates that these
systems are older (probably the supermassive compact galaxies seen at high redshifts) and have less recent star formation than either satellite or field elliptical galaxies, with growth in size and mass predominantly due to dry minor mergers \citep{Bernardi:2009p21529,BoylanKolchin:2006p21539}
If the progenitors of these BCGs are dwarfs, then it seems a challenge
to understand how metal-rich populations can be built up through the
successive mergers of metal poor populations.

There are some additional caveats to bear in mind when comparing Figure~\ref{fig:LvsRBCG} with
previous work. First we do not probe very massive clusters. Only 731 galaxies are in halos with masses greater than $\log(M_{halo}/ M_\odot)>14.0$. Thus we may be missing the very massive BCG galaxies present in the C4 or MaxBCG catalogs. Our redshift restrictions (z$<$0.1) may also cause us to miss more massive BCGs. Secondly, we have not accounted for sky over-estimates which adversely affect size measurements of massive clustered galaxies. It should be noted that \cite{Guo:2009p19543} fit their sample of galaxies with GALFIT \citep{Peng:2002p2528} to account for sky over-estimation while \cite{Weinmann:2009p19068} did not but found consistent results. 
We performed a test of our results using the corrections for de~Vaucouleurs sizes and magnitudes prescribed by \cite{Hyde:2008p12893}. Although the distribution for all ellipticals shifts to larger sizes and brighter magnitudes, the trends remain the same.
Thirdly, there may be a discrepancy in the cluster detection algorithm and BCG identification. The C4 and MaxBCG catalogs use the red-sequence technique to initially identify clusters whereas the Yang group catalog uses a modified friends-of-friends algorithm. The C4 and MaxBCG clusters will therefore miss blue clusters. In addition, recent work by \cite{vonderLinden:2007p21830} has actually found that not all galaxies identified as central galaxies in the C4 catalog are centrals. This effect cannot be ruled out for the \cite{Yang:2007p19054} group catalog.

Given these many caveats, we conclude that we have relatively little to say about the role played by BCG galaxies
in conditioning the luminosity-size relationships explored in this paper. However, they clearly have
a part to play, and the presence of BCGs probably goes a long way toward explaining the
large sizes of early-type galaxies at the 90th percentile (the final column) in Table~\ref{tab:relations}. 
We can confirm the tightness of the luminosity size relation of BCGs in comparison to satellite or field ellipticals.
We also find no difference between late stage satellite and field galaxies.

\section{Which size is appropriate?}
\label{sec:size}

Giuricin et al. (1988) first drew attention to the surprising
fact that the luminosity-radius relation provides the tightest of all
correlations observed among the photometric parameters that can be
used to characterize galaxies. This conclusion was subsequently
strengthened and confirmed by Gavazzi et al. (1996). More recently
van den Bergh (2008) has pointed out that the tightness of the
luminosity-radius relation among early-type galaxies is difficult
to reconcile with scenarios in which such galaxies are formed
via chaotic hierarchical mergers. In the present work we have used
the Petrosian (1976) outer radii R(90) of galaxies that enclose 90\% of the galaxian flux. 
Similarly van den Bergh (2008) used the radii of galaxies (corrected for reddening)
at the faint outer isophote having B = 25.0 mag arcsec$^{-2}$. We now
inquire whether a similarly tight correlation is observed between
the luminosities of galaxies and their effective Petrosian radii
R(50). The relevant observations are collected in Table~\ref{tab:relationsR(50)}. The
data in this table show strong correlations between galaxy luminosity and half-light radius. 
However, in all cases, the dispersion in the relations using R(50) is seen to be slightly larger 
than that which employs R(90). Perhaps more surprising is the lack of dependence of the slope 
of the luminosity-size relation on the morphology of early type galaxies (E, S0 and Sa galaxies) with R(90) 
whereas R(50) does distinguish between different early type galaxies.

We repeated our analysis with R(50), corrected as specified by \cite{Graham:2005p19601} to recover the
effective radii and total luminosity. Results are shown in Table~\ref{tab:Reffrelations} and are consistent with those found with R(50) and R(90).
From Table~\ref{tab:Devrelations}, using the SDSS provided de Vaucouleurs radii (corrected as specified by \cite{Hyde:2008p12893}) leads to a shallower slope but a larger scatter than R(90) for early type galaxies. The trends remain consistent. In other words the outermost radii of galaxies correlates most strongly with luminosity. This result suggests that factors such as mergers and star formation history affect the internal distribution of galaxy light more strongly than they do the overall size of these objects (see Appendix A).

\section{DISCUSSION}
\label{sec:discussion}

One of the most unexpected conclusions of the present work is
that elliptical galaxies (in all environments) exhibit such a tight
correlation between their sizes and their luminosities. This result is
unexpected if, as proposed by \cite{Toomre:1977p8354}, elliptical galaxies
all formed from mergers of spirals. Such mergers would be
expected to be chaotic and are therefore expected to feed into a wide
range of final states. The fact that E galaxies are actually observed to
exhibit a tight correlation between their sizes and their luminosities
suggests that they actually arrived at their present state along rather
similar evolutionary tracks. In other words it appears as if ellipticals
have both simple morphologies and similar evolutionary histories.
This conclusion was also reached by
\cite{Totani:2009p21552}, who noted that ``the evolution
of early type galaxies must occur ubiquitously, rather than stochastically.''

Recent N-body simulations and semi-analytic models \citep{Nipoti:2009p22761} are able to 
account for the small scatter in the Fundamental Plane specifically by forcing the formation of 
$\thicksim$ 50\% of the mass of present elliptical galaxies at
z$>$1 and allowing subsequent growth via dry mergers so as to account for the tight spread in ages
observed locally \cite{Thomas:2005p22786,Graves:2009p22588,Graves:2009p22581}. 
Simulations have shown that once the FP is established 
it is relatively insensitive to (a few) episodes of dry major mergers.
However the projections of the scaling relations, especially the scatter, are still expected to 
depend on merger history, orientation,  and mass ratios of the dry merger \cite{BoylanKolchin:2006p21539,Robertson:2006p15049}. 
Contrary to our results, this should be dependent on environment as there is  definite growth of 
elliptical galaxies occurring in dense environments.

The tight luminosity-size relationship further complicates
the now well-known problem of explaining the `red nuggets'
seen at high redshift (highly compact early-type galaxies).
The relevant observations are well described  in
\cite{vanDokkum:2008p2119}, \cite{Toft:2009p21778}, \cite{Damjanov:2008p18138}, 
 \cite{Collins:2009p22133} and Ê\cite{vanDokkum:2009p21522}. The
central difficulty is that the high-redshift systems are
anomalously compact for their masses (or luminosities),
but major mergers grow galaxies along the fundamental
plane, so it is difficult to reconcile the extreme size growth
needed with episodes of dry merging, particularly because the
galaxies seen are massive to begin with (but see \cite{Valentinuzzi:2009p19676}).  

Recently, \cite{Bernardi:2009p21529} (and others) have suggested that BCGs 
could be the low redshift equivalent to high redshift red-nuggets. Around
90\% of the mass of these systems have been established by z$\sim 1$ 
and the evolution required is a three-fold increase in size from high-redshifts
to the present epoch. 
Simulations predict that it is possible for a series of many minor dry mergers to successfully 
grow elliptical galaxy sizes while decreasing their central concentration \citep{Naab:2009p18639,Hopkins:2009p19544,Hopkins:2009p21531}. 
However, in simulations \cite{Nipoti:2009p22599} have shown that it is impossible to reproduce 
both the slope and the scatter of the size-mass relation by dry major or minor mergers while accounting for the red nuggets. In addition BCGs exhibit the smallest scatter in the size-luminosity plane out of all early type galaxies even though their sizes are evolving more strongly than normal ellipticals \citep{Bernardi:2009p21529}.

\cite{Kormendy:2009p21817} certainly makes a strong case for the assumption that the internal structure of early-type galaxies ({\em i.e.}  the presence or absence of cores) depends on merger history. 
The curvature in the scaling relation for elliptical galaxies (noted by other authors) due to BCGs 
could also be a signature of minor mergers \citep{BoylanKolchin:2006p21539}, although
it is worth emphasizing the important caveat that
\cite{Masjedi:2007p262} find that major mergers are the dominant mode
of growth at $z<$0.3, and that at higher redshifts minor mergers
are difficult to detect directly.
In any case, it seems
a huge challenge to try to explain the build-up of a metal rich giant population through
successive minor dry mergers of metal-poor dwarf populations. This mismatch in
metallicities 
seems to us to
be an absolutely fundamental problem with this basic picture, 
and this
difficulty is only 
compounded by our
observation of a 
very small dispersion in the luminosity-size relation for BCGs. 

It seems hard to see how any stochastic process could lead to
the very tight luminosity-size relationship reported here, so on the whole our results
seem to favor the notion that mergers have not
greatly affected the luminosity-size relation of elliptical galaxies.
Since mergers are clearly happening \citep{Masjedi:2007p262},
but they do not appear to be having a great effect
on the luminosity-size relationship, perhaps the
mergers are only superficial signatures masking
some more important underlying process
(rearranging the deck chairs on the sinking Titanic). 
In any case, our observations suggest that
concentration and age are responsible for
much of the variance in the luminosity-size
relation, and perhaps this provides a useful constraint on models.

In spite of de-emphasizing the role of mergers in the previous
paragraph, it would be a mistake to conclude from
our work that the luminosity-size relationship is independent
of environment. In fact, a striking feature of Figure~\ref{fig:LvsRperTTyperho}  is that large
elliptical galaxies strongly prefer dense environments.
This point is seen particularly clearly by inspection of
the range in sizes spanned by the 10\% to 90\% size
quantiles tabulated in Table~\ref{tab:relations}. We have already noted
that much of this effect may find its origin in the presence of
BCGs in our sample, and the physics of size growth in these
objects may well differ from that of elliptical galaxies in
sparse environments. At the same time, the
absence of
under-luminous galaxies  in dense regions
is probably the natural extension of the effect seen
by \cite{Coenda:2009p18836} in their cluster sample,
and follows the general trends noted by 
 \cite{Blanton:2003p2277,Blanton:2005p2271}
and \cite{Hogg:2004p294}, although these authors did
not study the morphologies of the galaxies in any detail. The central
point emerging from our study is that even though the
linear fits to the luminosity-size data for
E is unchanged with environment, 
galaxies with sizes greater than $\log(R_{90}/{\rm kpc}) > 1.2$  are abundant
in dense regions while being essentially absent
from sparse regions in the field. So even though
the {\em form} of the luminosity-size relationship seems
independent of environment, the {\em range spanned} by
galaxies described by the relation seems to be a strong function of
environment. This result would seem to follow
naturally from the work of \cite{Zucca:2009p21822},
who find, as we also do, that the characteristic
magnitude of galaxies is fainter in under-dense regions than
it is in over-dense regions.

It is interesting to note that our analysis includes
255 early-type
galaxies in the blue cloud. While this is only a small
fraction (8\%) of the total early-type sample,
these are exceptionally interesting objects.
Careful re-inspection of the images suggests that
some small fraction of these could be misclassifications
of later-type galaxies, but the great majority of the
classifications seem secure (see \cite{Nair:2010p22790}).
These blue cloud early types could be ellipticals which have undergone 
a wet major merger recently, which has led to the reformation of a disk \citep{Wei:2009p21510},
although
it is interesting to note that blue cloud early types show a shallow
slope on the luminosity-size relationship diagram and 
hence do not behave like disk galaxies. It is not clear how to 
reconcile this shallower slope with the claim by
 \cite{Wei:2009p21510} that the presence of E and S0 galaxies on the blue cloud suggests that wet major merger processes could lead to the reformation of a disk. (The dispersion for these systems are larger and environmentally dependent.)
Perhaps wet mergers do not affect the luminosity-size relationship
in the same way as dry mergers affect the relationship.

The absence of luminous (or large) galaxies in sparse
regions is also notable amongst the late-type population.
Figure~\ref{fig:LvsRperTTyperho}  shows a striking deficiency of faint late-type spirals
in high-density regions. Very few Sb or Sc spirals, in regions of above-average
density, have log $Lg/L_{\odot}$ $<$ 9.5.
In other words the dispersion
in the luminosity-radius relation of galaxies increases steeply
towards later morphological types. In the case of spirals of type
Sc this increased dispersion is, at least in part, due to a strong
population of faint objects that fall below the line defined by
the fiducial relationships shown in Table~\ref{tab:relations} (in other words,
the dispersion is skewed to small systems).

Figure~\ref{fig:LvsRperTType} and Table 1 also provide some insight into the relationship between
elliptical galaxies and S0 systems. The basic conclusion from
\cite{vandenBergh:2004p9629} 
that lenticular galaxies are (as a class) less luminous that either elliptical
galaxies or spirals of type Sa is strongly reinforced
by these data. Furthermore, for S0 galaxies, the dispersion
of the data around the fiducial relationship is found to be
0.16 dex. This value is greater than that for Sa galaxies but only at the 1$\sigma$ level,
although it is significantly larger than that for elliptical galaxies (over 5$\sigma$ significant).
These results seem consistent with the conclusion by \cite{vandenBergh:2009p21551} that 
S0 galaxies are not intermediate between E and Sa galaxies as \cite{Hubble:1936p14278} had proposed. 
Inspection of the data plotted in
Figure~\ref{fig:LvsRperTType}  suggests that the increased dispersion of S0 galaxies is due to the
addition of a population of objects that are either exceptionally luminous for their 
radius, or exceptionally small for their luminosity.

\section{Conclusions}
\label{sec:conclusion}

The present investigation is based on a sample of 12150 galaxies that
are contained in the Sloan Digital Sky Survey \citep{York:2000p3192}. Visual
morphological classifications by \cite{Nair:2010p22790}, and environmental
information from \cite{Blanton:2005p79}, \cite{Baldry:2006p103} and \cite{Yang:2007p19054}  were employed to study the dependence
of the luminosity-radius relation of these objects on both their
morphological type and environmental density. The
main conclusions of this work are: 
\begin{enumerate} 
\item The luminosity-size relation
of elliptical galaxies is well represented by a simple power law.
The scatter about this relationship is very small:
for elliptical galaxies, the dispersion is 
only 0.14 dex (0.35 mag), making this relationship
as tight as that defined by the fundamental plane,
even though it is based on purely photometric parameters.
\item The scatter about this luminosity-size relation is due to variations
in both galaxy central concentration and star-formation history.
Increasing central concentration
has the effect of making modest changes
in the slopes and intercepts of the fiducial relations.
Although for all subsets of early-type galaxies
the scatter remains small,
the scatter in the luminosity-size relationship
is $\sim50$\% larger for a small fraction ($<10\%$) of
early-type galaxies in the blue cloud than on the red sequence.
\item The slope and dispersion about the luminosity-size relationship
increases as one proceeds along the Hubble sequence $E \Rightarrow {\rm Sa} \Rightarrow {\rm Sb}  \Rightarrow {\rm Sc}$.
\item Elliptical galaxies are presumed to have formed by the chaotic mergers of either
smaller ellipticals or of disk-like ancestral objects, but it is difficult to see how 
such a process could result in a very tight luminosity-size relationship.
\item For galaxies of a given Hubble type,
the slope of the luminosity-size relationship seems
independent of environment. However
the range spanned by the luminosities (or sizes) of
that class of galaxy seems to be a strong function of
environment.
\item The de Vaucouleurs radius, Rdev, and Petrosian half light radius, R(50),
are more sensitive to changes in central concentration and hence merger
history than R(90). Thus it appears mergers only influence the internal structure
of galaxies and not their global luminosity-size relation.

\end{enumerate}
It is
not clear why objects that might have assembled in such very different
ways, from differing ancestral objects, should have had evolutionary
tracks that converged to  show small dispersions around 
simple power-law forms for their
luminosity-size relationships. Perhaps elliptical
galaxies  started out very differently from each other at $z\sim3$,
but their morphologies converged to similar configurations
as they expanded between redshifts $z = 3$ and $z = 0$ \citep{Johansson:2009p18681}.

\acknowledgments
\noindent{\em Acknowledgments}

PN and RGA thank Ivan Baldry for a helpful discussion about possible selection
effects affecting the luminosity-radius relations for galaxies in
differing environments. PN would also like to thank Gianni Zamorani, Luca Ciotti and Carlo Nipoti for helpful discussions.
SvdB would like to thank Leo Girardi, David Hogg, Igor Karachentsev, and John Kormendy
 for helpful exchanges of e-mails.

Funding for the SDSS and SDSS-II has been provided by the Alfred P. Sloan Foundation, the Participating Institutions, the National Science Foundation, the U.S. Department of Energy, the National Aeronautics and Space Administration, the Japanese Monbukagakusho, the Max Planck Society, and the Higher Education Funding Council for England. The SDSS Web Site is http://www.sdss.org/.

The SDSS is managed by the Astrophysical Research Consortium for the Participating Institutions. The Participating Institutions are the American Museum of Natural History, Astrophysical Institute Potsdam, University of Basel, University of Cambridge, Case Western Reserve University, University of Chicago, Drexel University, Fermilab, the Institute for Advanced Study, the Japan Participation Group, Johns Hopkins University, the Joint Institute for Nuclear Astrophysics, the Kavli Institute for Particle Astrophysics and Cosmology, the Korean Scientist Group, the Chinese Academy of Sciences (LAMOST), Los Alamos National Laboratory, the Max-Planck-Institute for Astronomy (MPIA), the Max-Planck-Institute for Astrophysics (MPA), New Mexico State University, Ohio State University, University of Pittsburgh, University of Portsmouth, Princeton University, the United States Naval Observatory, and the University of Washington.

\bigskip
\appendix
\bigskip

\section{A. Biases in measurements of size?}
\label{sec:seeing}

Three issues need to be kept in mind while using SDSS sizes and luminosities:\\

\noindent 1. SDSS has a known problem of sky over-estimation near bright galaxies especially in dense regions. We have partially compensated for this by selecting a clean sample of galaxies with no overlapping companions. In addition, we ran our analysis excluding objects with r-band model magnitudes r$<$14 where the effects of sky overestimation are severe. Changes in the fits are small and within the error bars quoted. Our overall results and conclusions remain unaffected.\\

\noindent 2. Petrosian sizes have not been corrected for seeing.  

\begin{figure*}[t!]
\unitlength1cm
\begin{minipage}[t]{4.0cm}
\rotatebox{0}{\resizebox{8cm}{5cm}{\includegraphics{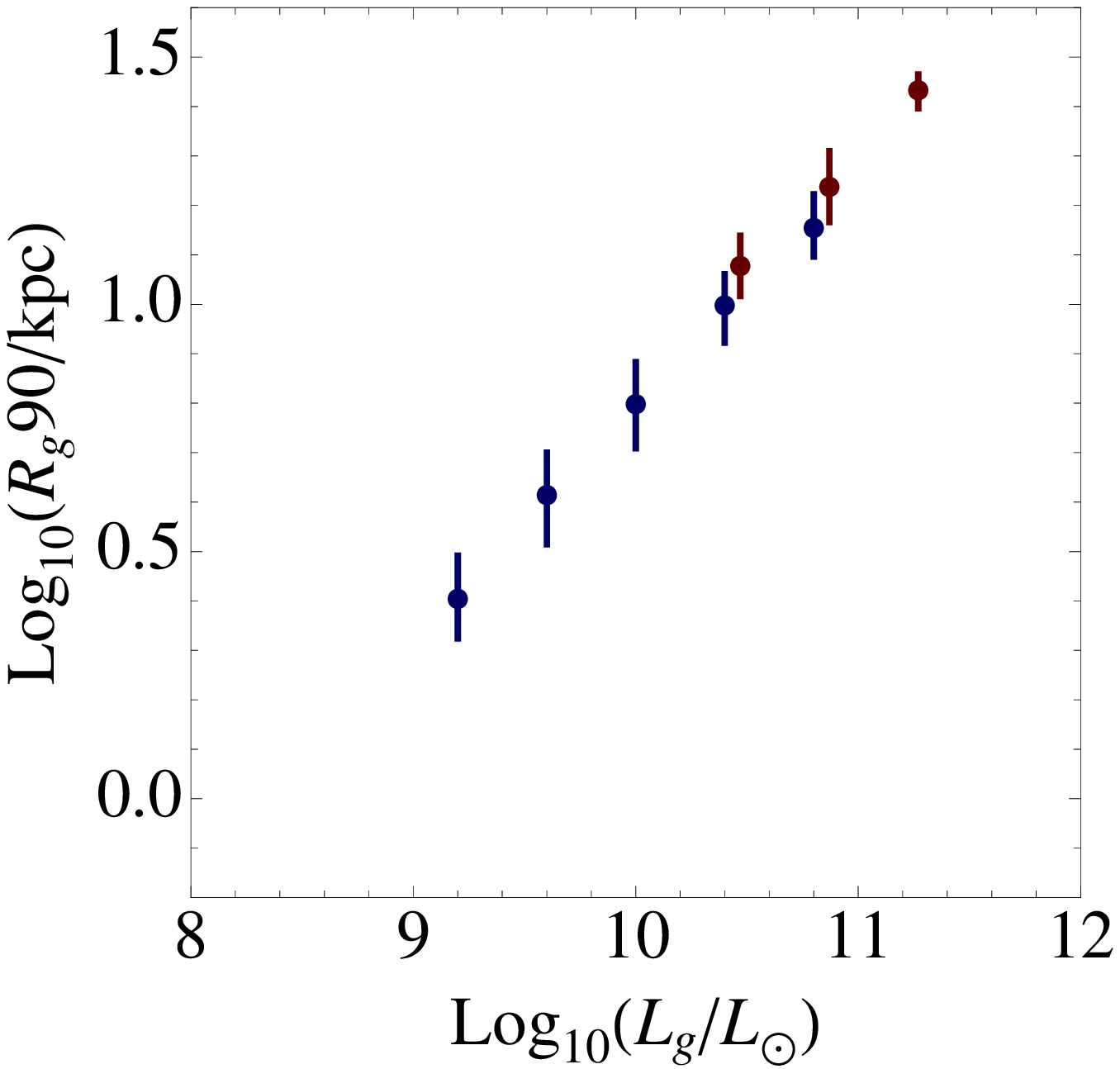}}}
\rotatebox{0}{\resizebox{8cm}{5cm}{\includegraphics{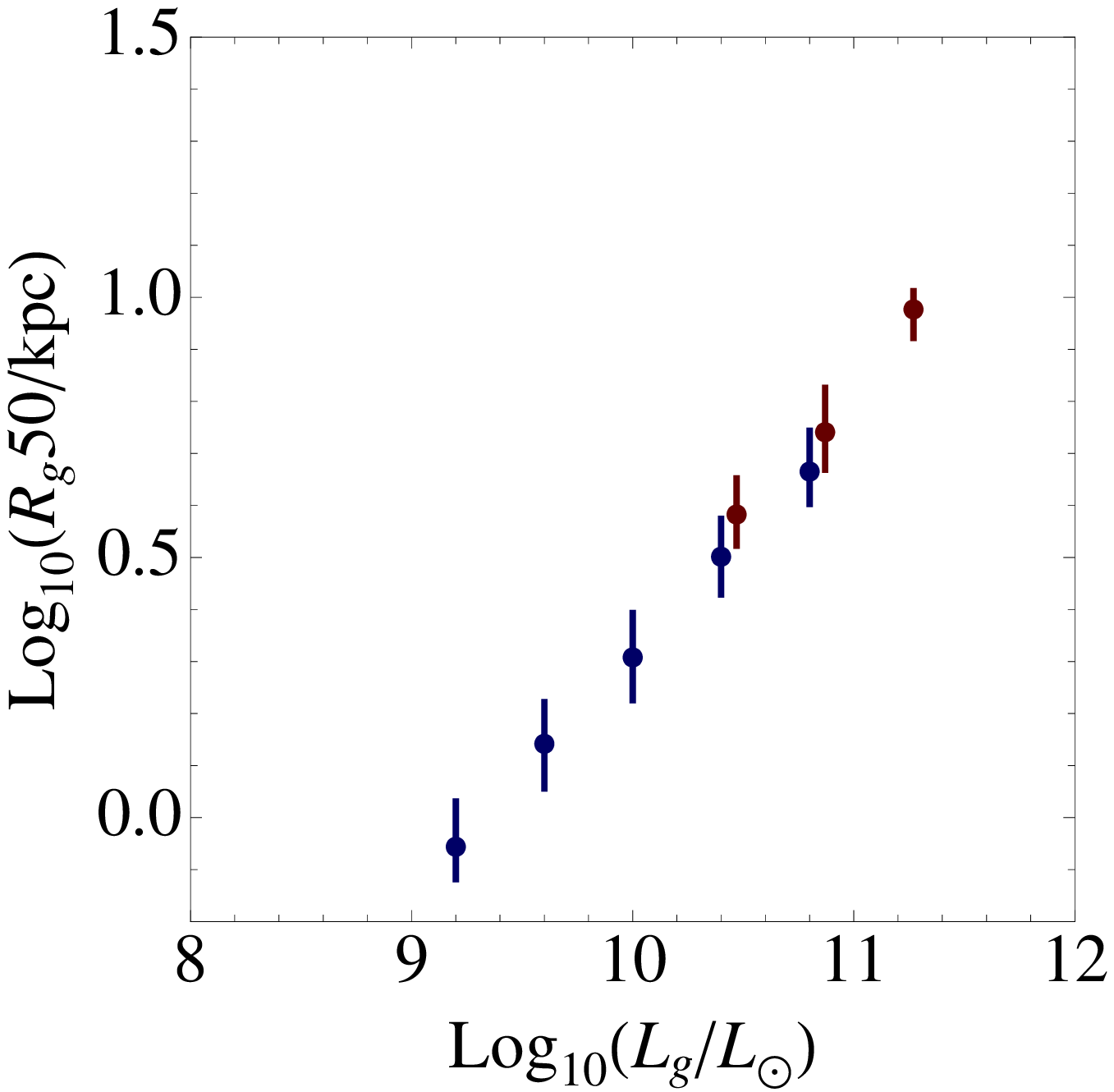}}}
\end{minipage}
\hspace{3.0cm}
\begin{minipage}[t]{4.0cm}
\rotatebox{0}{\resizebox{8cm}{5cm}{\includegraphics{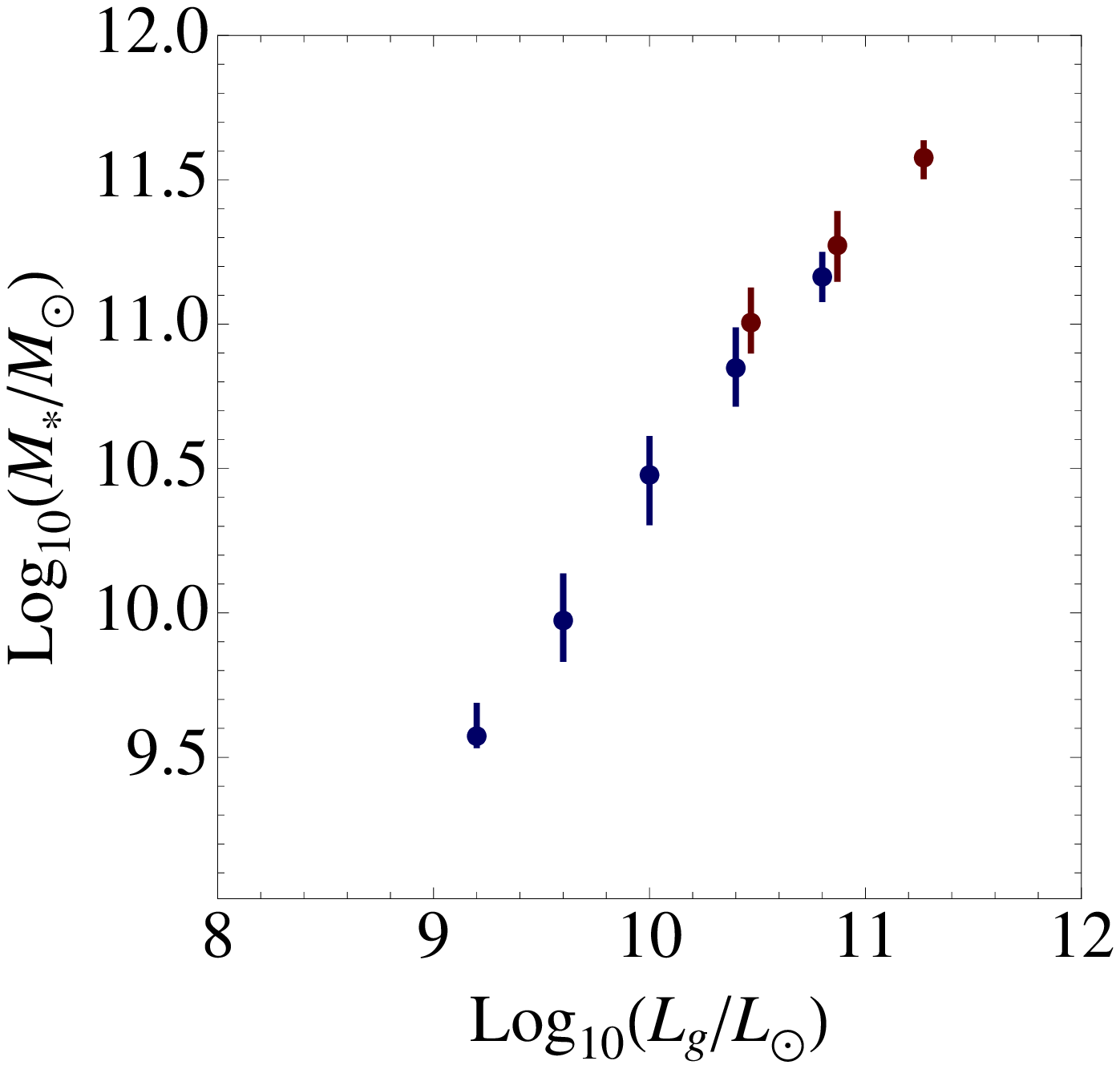}}}
\rotatebox{0}{\resizebox{8cm}{5cm}{\includegraphics{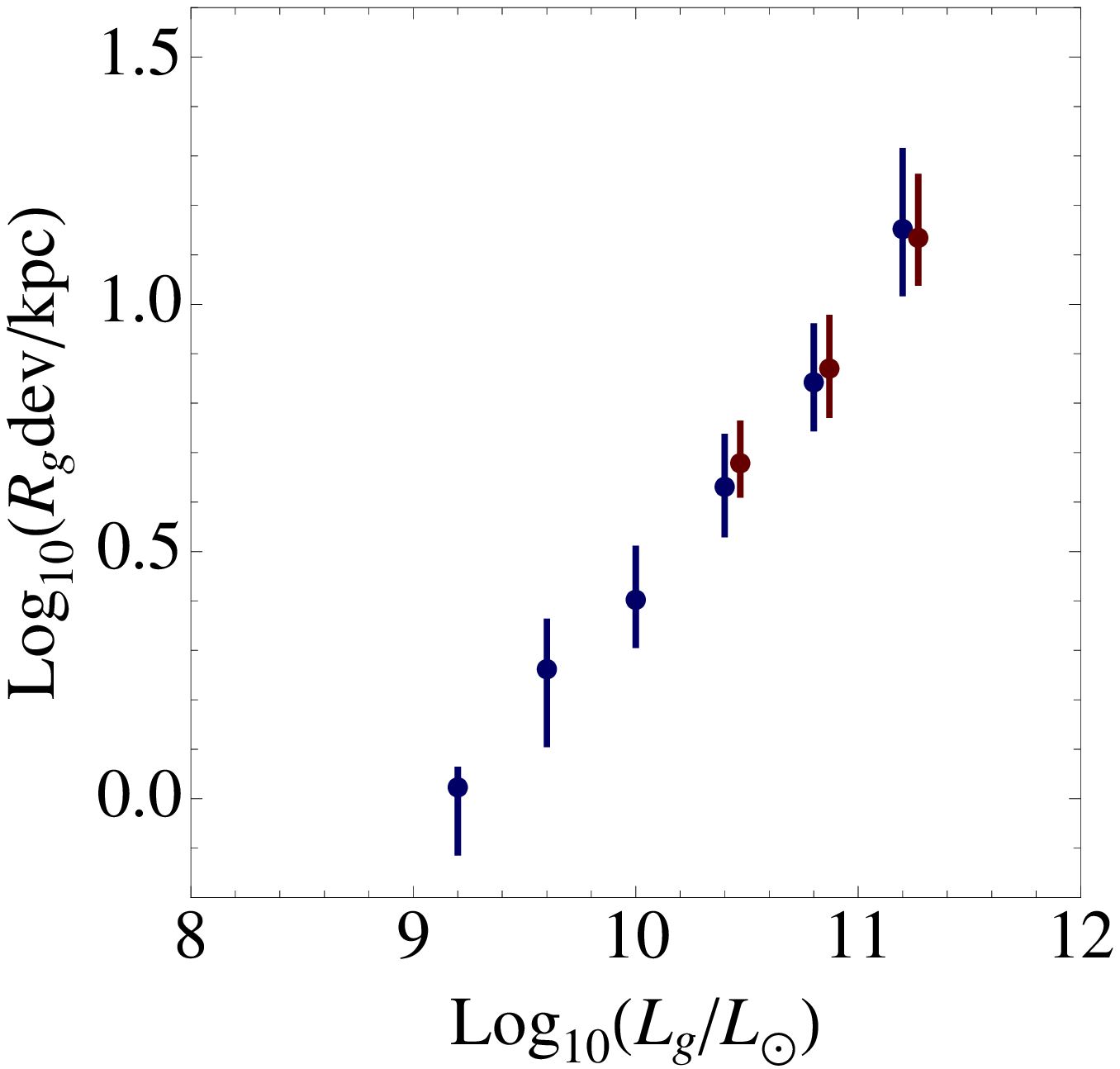}}}
\end{minipage}
\caption[Plot]{\label{fig:SeeingDependence}  (a) g-band Petrosian R(90) radius in kpc versus g-band luminosity for ellipitical galaxies, (b) stellar mass versus g-band luminosity for ellipitical galaxies, (c) g-band Petrosian R(50) radius in kpc versus g-band luminosity for ellipitical galaxies and (d) g-band de Vaucouleurs radius in kpc versus g-band luminosity (corrected for sky subtraction as per \cite{Hyde:2008p12893}) for ellipitical galaxies. The red points show galaxies at redshifts above 0.05 while blue points show galaxies below it. The red points are offset for clarity.
  }
\end{figure*}

\cite{Hyde:2008p12893} use the SDSS seeing corrected de Vaucouleurs size (RdeV) and magnitude
as they find the SDSS Petrosian quantities (which are not seeing corrected) show a slight bias 
where higher redshift early type galaxies are larger than lower redshift early type galaxies at similar luminosities. 
As we probe a smaller redshift range z$<$0.1 than Hyde et al. (z$<$0.3), we do not expect seeing to be important in measures of R(90). 
To illustrate this, Figure~\ref{fig:SeeingDependence}(a) shows the median  g-band Petrosian R(90) size in bins of luminosity for ellipticals galaxies. The error bars show the 25th and 75th quantiles. The blue points represent galaxies below a redshift of 0.05 while the red points show galaxies above z of 0.05. We find galaxies in the higher redshift bin are slightly larger at any given luminosity which could imply a seeing bias. However, Figure~\ref{fig:SeeingDependence}(b) shows the median stellar mass in bins of luminosity where we find a comparable redshift dependence of mass on luminosity. This is because the higher redshift bin has more massive galaxies (volume effect) and fewer lower mass systems due to  our apparent magnitude cut (g$<$16). Thus the offset of size measures at similar luminosities is because of different mass selections and not due to seeing. Surprisingly Figure~ \ref{fig:SeeingDependence}(c) shows the Petrosian R(50) size has a comparable redshift dependent offset with luminosity. Figure~ \ref{fig:SeeingDependence} (d) shows the g-band de Vaucouleurs size (Rdev) versus luminosity, both corrected for over-estimation of sky as specified by \cite{Hyde:2008p12893}. The redshift dependent offset in the size measures are smaller but the scatter in each redshift bin is larger than the scatter in R(90) or R(50). This is especially noticeable in the highest luminosity bins where the scatter in R(90) is much smaller than Rdev.\\

\noindent 3. The choice of a size metric influences the scaling relations.

\begin{figure*}[b!]
\unitlength1cm
\begin{minipage}[t]{4.0cm}
\rotatebox{0}{\resizebox{8cm}{5cm}{\includegraphics{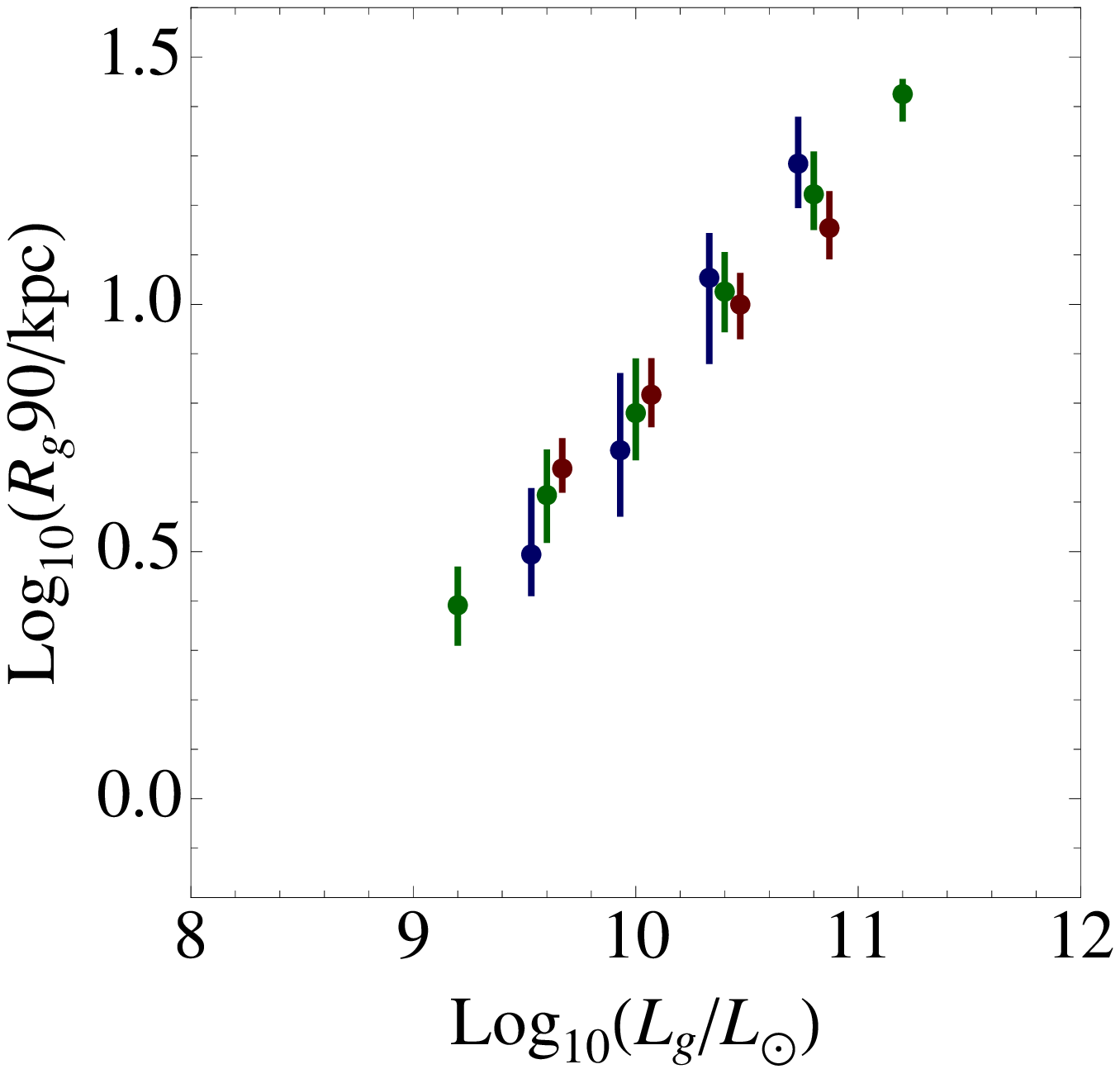}}}
\rotatebox{0}{\resizebox{8cm}{5cm}{\includegraphics{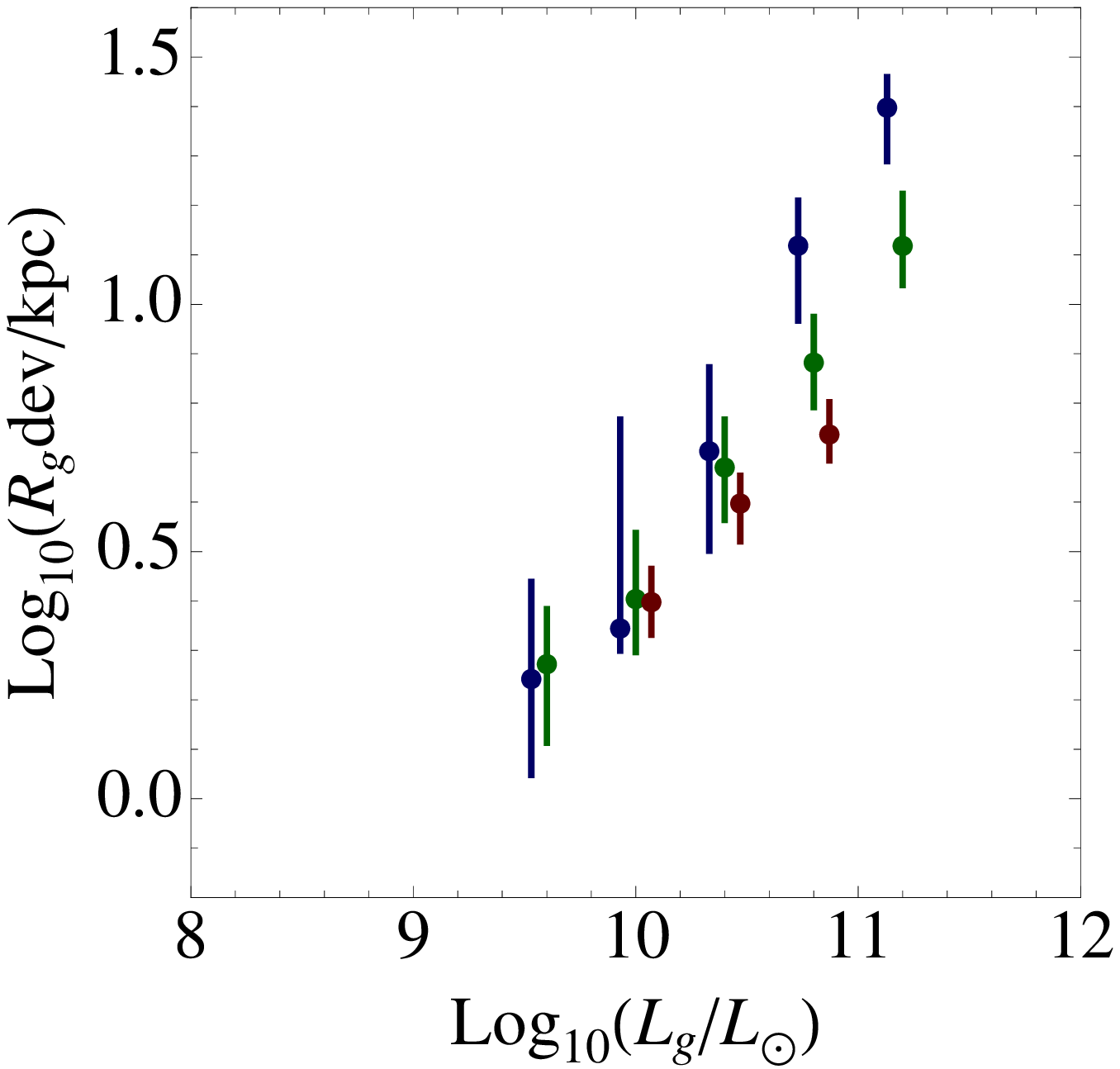}}}
\end{minipage}
\hspace{3.0cm}
\begin{minipage}[t]{4.0cm}
\rotatebox{0}{\resizebox{8cm}{5cm}{\includegraphics{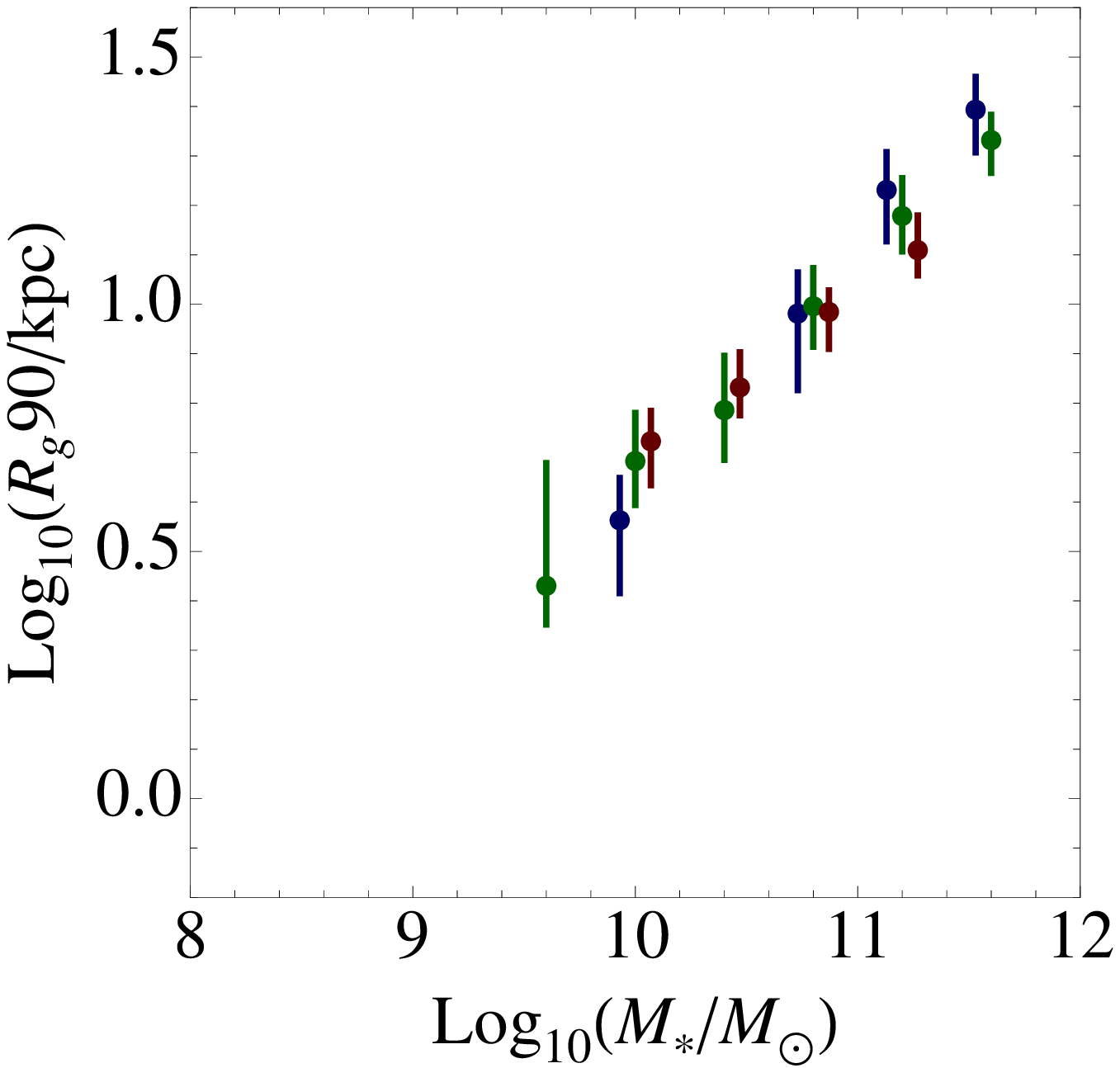}}}
\rotatebox{0}{\resizebox{8cm}{5cm}{\includegraphics{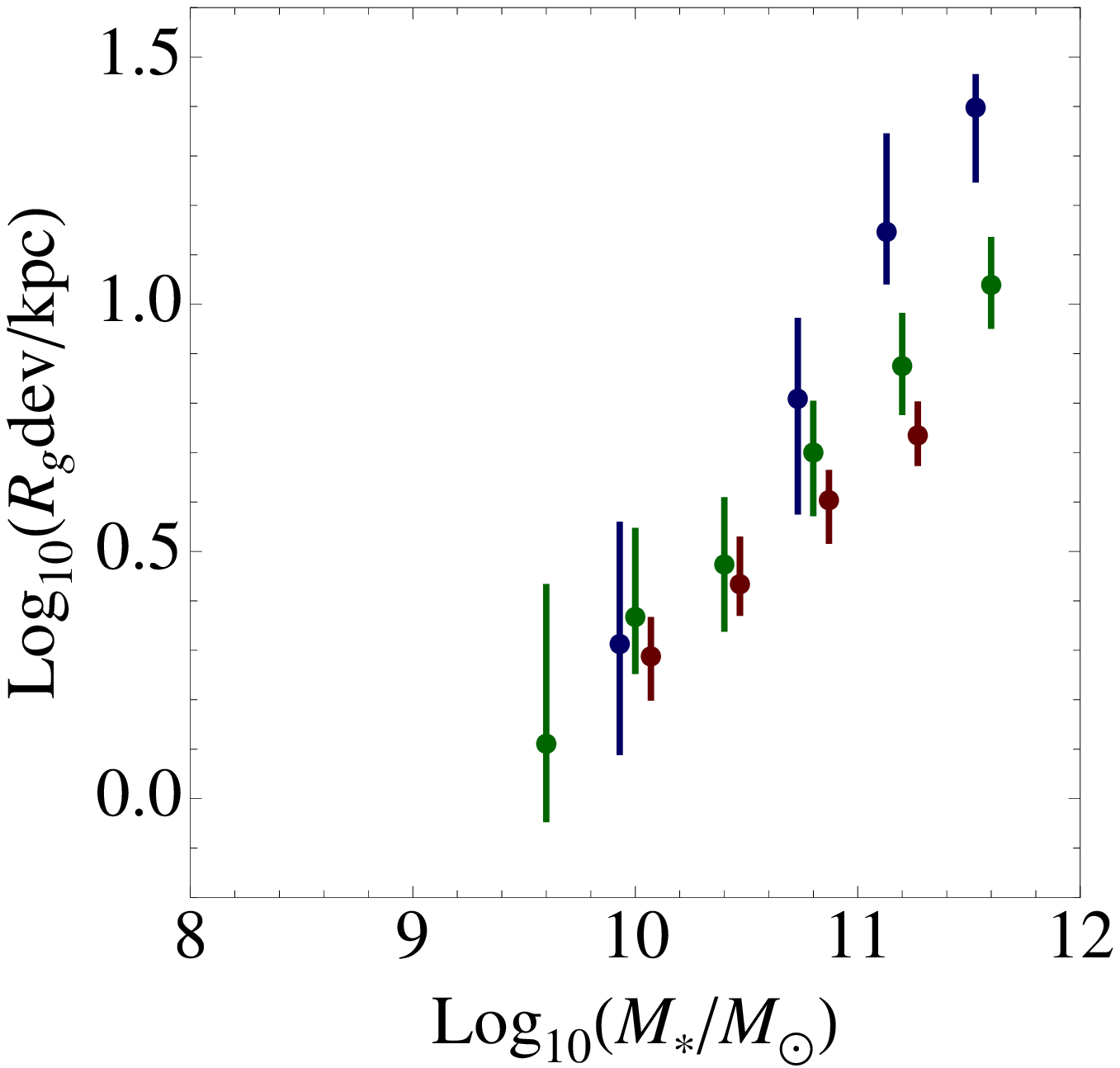}}}
\end{minipage}
\caption[Plot]{\label{fig:ConcentrationDependence}  Top: g-band Petrosian R(90) radii in kpc versus (a) luminosity and (b) stellar mass. Bottom: $R_{dev}$ versus (c) Dev luminosity and (d) stellar mass. Points are keyed to central concentration. The most concentrated galaxies  C$>$0.5 are shown in red, followed by green 0.4$<$C$<$0.5  with the least concentrated galaxies shown in blue (C$<$0.4). de Vaucouleurs measure of size is very dependent on the concentration of the galaxy, especially at high masses/luminosities. The red points are offset for clarity.
  }
\end{figure*}

Figure~\ref{fig:ConcentrationDependence}  shows the dependence of Petrosian R(90) (top) and de Vaucouleurs radii (bottom) on luminosity (left) and stellar mass (right) as a function of concentration. The most concentrated galaxies are shown in red (C$>$0.5), followed by green (0.4$<$C$<$0.5),  with the least concentrated galaxies shown in blue (C$<$0.4). The de Vaucouleurs measure of size is strongly dependent on the concentration of the galaxy, especially at high masses/luminosities whereas the dependence of R(90) on concentration is mild.  Thus Rdev is more sensitive to the internal structure of early type galaxies. The curvature seen in high density environments may be related to physical processes which change the central concentration of galaxies while not changing R(90).  In other words, the reason for differing results among various authors regarding curvature in scaling relations may not be because of seeing dependent measures of size but on the effect of concentration on those measures. Thus we would recommend using R(90) which is relatively insensitive to both seeing and central concentration to the de Vaucouleurs size measure. It should be noted that our results for elliptical galaxies do not change when using Rdev. Table~\ref{tab:Devrelations} shows fits for elliptical galaxies using de Vaucouleurs size and magnitude (sky corrected). We find the slopes are the same in low and high density environments with a smaller scatter in high density environments. BCG galaxies in clusters show a shallower luminosity-size relation (steeper size-luminosity relation) than satellite galaxies.

\section{B. Comparison of environment estimates}
\label{sec:env}

\begin{figure*}[t!]
\unitlength1cm
\hspace{1cm}
\rotatebox{0}{\resizebox{16cm}{18.5cm}{\includegraphics{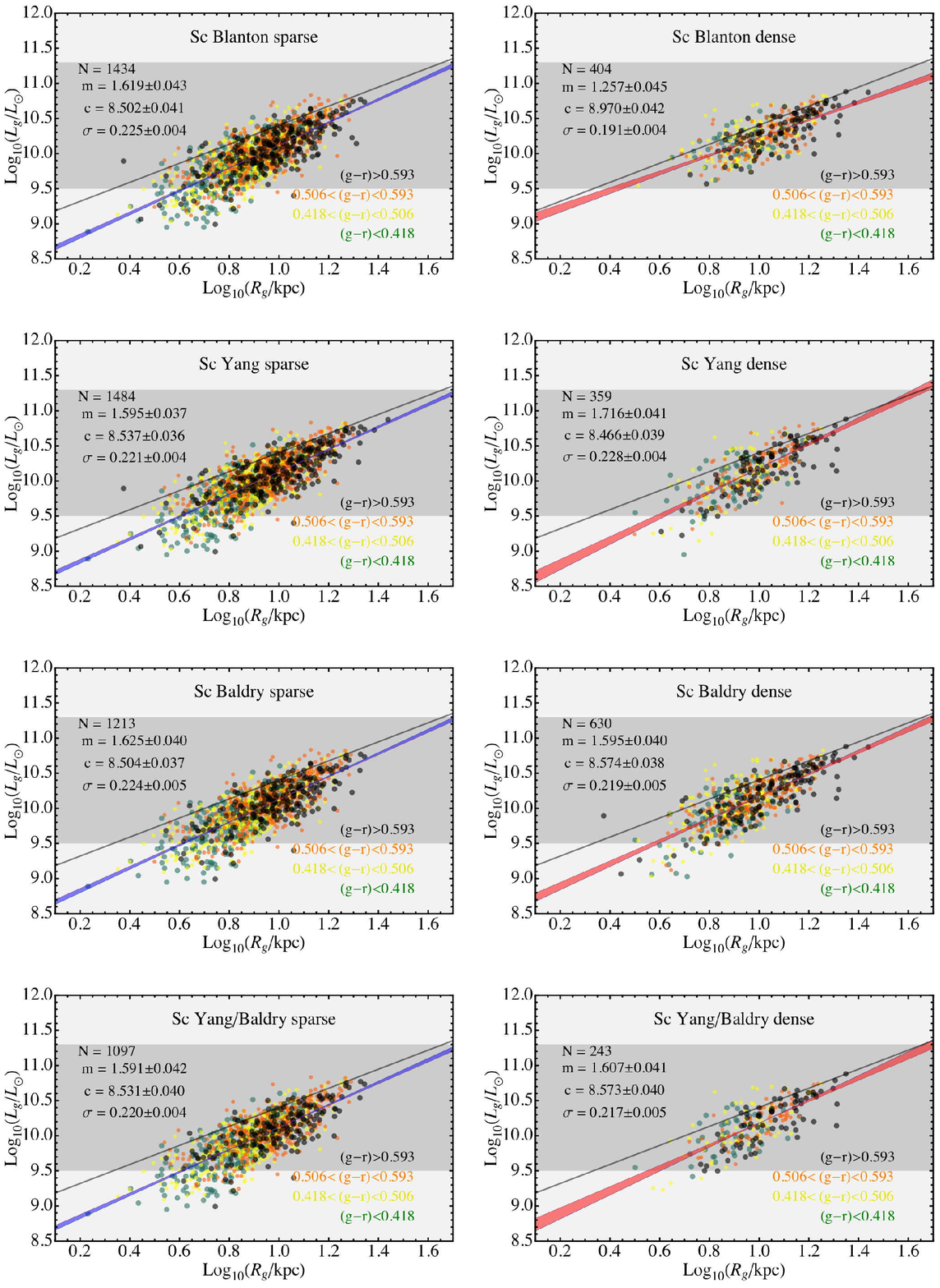}}}
\caption[Plot]{\label{fig:LvsRScEnc} g-band Luminosity versus g-band Petrosian R(90) radius in kpc in low (left) and high (right) density environments for Sc galaxies only. The top three rows show environments defined by Blanton et al 2005, Yang et al. 2007 and Baldry et al. 2006. The bottom panel shows L vs. R for galaxies defined using both Yang and Baldry estimates of environmental density. The points are keyed to $g-r$ color quartiles. Black points are the reddest galaxies in each panel, followed by orange, then yellow with green points being the bluest galaxies.
}
\end{figure*}

\begin{figure}[t!]
\unitlength1cm
\rotatebox{0}{\resizebox{10cm}{6cm}{\includegraphics{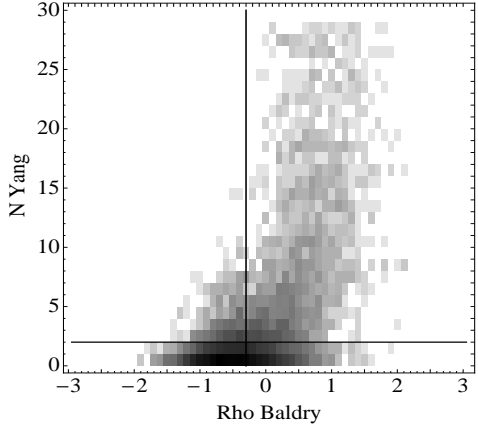}}}
\caption[Plot]{\label{fig:EnvComparison} Density plot showing comparison of \cite{Baldry:2006p103} nearest neighbor environmental estimate versus \cite{Yang:2007p19054}  group number estimate. The vertical line show the median Baldry density estimate (Log $\Sigma$ = -0.32) while the horizontal line (N=2) is our demarcation for galaxies in the field versus groups/clusters. N$\le$2 are under dense regions. N$>$2  are over-dense regions. For clarity, we do not show groups with more than 30 members in this plot.
}
\end{figure}

Figure~\ref{fig:LvsRScEnc} shows the g-band luminosity of Sc galaxies versus the g-band Petrosian R(90) radius in low and high density regions keyed to the color of the galaxy using the three environmental measures available. The first panel shows the distribution of Sc galaxies in low and high density regions as defined by \cite{Blanton:2005p79}, the second row as defined by \cite{Yang:2007p19054} and the third row as defined by \cite{Baldry:2006p103}. The last row is defined using both Yang and Baldry measures of environment. The black line in each panel shows the relation for elliptical galaxies. This figure shows that many small Sc galaxies in clusters are missed when using the Blanton et al. measure of environment, which leads to a spurious apparent evolution of the L vs R relation for Sc galaxies from the field to cluster. This bias is largest for late type galaxies and smallest for elliptical galaxies. We thus restrict ourselves to the Baldry and Yang measures of environment, compared in Figure~\ref{fig:EnvComparison}. We find for N$<$3, some galaxies are in over-dense regions implying satellites or companions were missed but most are in under-dense regions (i.e. field galaxies). 

\begin{deluxetable}{rcccccc} 
\tablecolumns{7} 
\tablewidth{0pc} 
\tabletypesize{\scriptsize} 
\tablecaption{luminosity-size R(90) relationships\label{tab:relations}}
\tablehead{ 
 \colhead{} & \colhead{} &  \colhead{} & \colhead{} & \colhead{} &
  \multicolumn{2}{c}{-- Interquantile Size Range -- } \\  
  \colhead{Class} & 
  \colhead{N} &
  \colhead{Slope} &
  \colhead{Intercept} &
  \colhead{Dispersion} &
  \colhead{10\% quantile} &
  \colhead{90\% quantile \vspace{0.1cm}} \\
  \colhead{} & 
  \colhead{} &
  \colhead{$[\log(L_\odot)]$} &
   \colhead{$\left[\log(L_\odot) \over \log({\rm kpc)}\right]$ \vspace{0.1cm}} &
   \colhead{$[\log(L_\odot)]$} &
  \colhead{$[\log({\rm kpc})]$} &
   \colhead{$[\log({\rm kpc})]$}
} 
\startdata
\\
 \cutinhead{\em Elliptical galaxies}    
     All & 3163 & $1.378 \pm 0.014$ & $9.026 \pm 0.016$ & $0.1425 \pm 0.0021$ & 0.65 & 1.36 \\   
     Sparse & 801 & $1.36 \pm 0.029$ & $9.032 \pm 0.031$ & $0.1619 \pm 0.0046$ & 0.56 & 1.24 \\   
     Rich & 1514 & $1.319 \pm 0.02$ & $9.101 \pm 0.023$ & $0.1271 \pm 0.0032$ & 0.74 & 1.4 \\   
     Rich  N$\ge$10 & 766 & $1.316 \pm 0.021$ & $9.108 \pm 0.025$ & $0.1224 \pm 0.0045$ & 0.7 & 1.43 \\   
     Rich  BCG N$\ge$10 & 374 & $1.132 \pm 0.038$ & $9.36 \pm 0.048$ & $0.09946 \pm 0.0041$ & 1. & 1.45 \\   
     Rich  satellite  N$\ge$10 & 392 & $1.263 \pm 0.04$ & $9.138 \pm 0.042$ & $0.1346 \pm 0.0066$ & 0.63 & 1.29 \\   
     Red sequence & 2908 & $1.403 \pm 0.014$ & $8.995 \pm 0.016$ & $0.1357 \pm 0.0023$ & 0.66 & 1.36 \\   
     Blue cloud & 255 & $1.185 \pm 0.057$ & $9.273 \pm 0.064$ & $0.1896 \pm 0.011$ & 0.55 & 1.31 \\    
     
   \cutinhead{\em S0 galaxies}    
     All & 1893 & $1.335 \pm 0.018$ & $9.043 \pm 0.017$ & $0.1636 \pm 0.0033$ & 0.56 & 1.19 \\   
     Sparse & 751 & $1.295 \pm 0.028$ & $9.086 \pm 0.026$ & $0.1597 \pm 0.0043$ & 0.55 & 1.16 \\   
     Rich & 562 & $1.352 \pm 0.032$ & $9.012 \pm 0.032$ & $0.1665 \pm 0.0055$ & 0.61 & 1.22 \\   
     Rich  N$\ge$10 & 306 & $1.321 \pm 0.043$ & $9.02 \pm 0.041$ & $0.1588 \pm 0.0067$ & 0.57 & 1.17 \\   
     Rich  BCG N$\ge$10 & 32 & $0.8086 \pm 0.11$ & $9.678 \pm 0.11$ & $0.09644 \pm 0.013$ & 0.78 & 1.26 \\   
     Rich  satellite  N$\ge$10 & 274 & $1.272 \pm 0.046$ & $9.048 \pm 0.042$ & $0.157 \pm 0.0067$ & 0.56 & 1.15 \\   
     Red sequence & 1520 & $1.315 \pm 0.019$ & $9.04 \pm 0.018$ & $0.1539 \pm 0.0031$ & 0.57 & 1.18 \\   
     Blue cloud & 373 & $1.316 \pm 0.048$ & $9.144 \pm 0.047$ & $0.175 \pm 0.0073$ & 0.54 & 1.21 \\    
     
   \cutinhead{\em Sa galaxies}    
     All & 1955 & $1.337 \pm 0.022$ & $9.008 \pm 0.023$ & $0.1604 \pm 0.0031$ & 0.63 & 1.25 \\   
     Sparse & 883 & $1.324 \pm 0.033$ & $9.018 \pm 0.033$ & $0.1536 \pm 0.0044$ & 0.63 & 1.23 \\   
     Rich & 484 & $1.322 \pm 0.04$ & $9.032 \pm 0.042$ & $0.1606 \pm 0.0058$ & 0.65 & 1.27 \\   
     Rich  N$\ge$10 & 237 & $1.291 \pm 0.059$ & $9.031 \pm 0.06$ & $0.1657 \pm 0.008$ & 0.65 & 1.23 \\   
     Rich  BCG N$\ge$10 & 22 & $1.167 \pm 0.25$ & $9.278 \pm 0.29$ & $0.1318 \pm 0.016$ & 0.94 & 1.31 \\   
     Rich  satellite  N$\ge$10 & 215 & $1.229 \pm 0.074$ & $9.079 \pm 0.075$ & $0.1648 \pm 0.0075$ & 0.64 & 1.21 \\   
     Red sequence & 1001 & $1.231 \pm 0.033$ & $9.078 \pm 0.033$ & $0.1619 \pm 0.0048$ & 0.61 & 1.23 \\   
     Blue cloud & 954 & $1.4 \pm 0.028$ & $8.975 \pm 0.029$ & $0.1513 \pm 0.0039$ & 0.68 & 1.26 \\    
     
   \cutinhead{\em Sb galaxies}    
     All & 2736 & $1.449 \pm 0.022$ & $8.854 \pm 0.024$ & $0.1719 \pm 0.0028$ & 0.73 & 1.28 \\   
     Sparse & 1281 & $1.424 \pm 0.027$ & $8.876 \pm 0.029$ & $0.1717 \pm 0.0039$ & 0.71 & 1.27 \\   
     Rich & 615 & $1.502 \pm 0.046$ & $8.812 \pm 0.05$ & $0.1684 \pm 0.0047$ & 0.78 & 1.32 \\   
     Rich  N$\ge$10 & 264 & $1.479 \pm 0.066$ & $8.8 \pm 0.07$ & $0.1682 \pm 0.0069$ & 0.77 & 1.3 \\   
     Rich  BCG N$\ge$10 & 36 & $0.7813 \pm 0.14$ & $9.738 \pm 0.16$ & $0.1198 \pm 0.017$ & 0.95 & 1.39 \\   
     Rich  satellite  N$\ge$10 & 228 & $1.429 \pm 0.077$ & $8.833 \pm 0.08$ & $0.1636 \pm 0.0067$ & 0.76 & 1.25 \\   
     Red sequence & 788 & $1.399 \pm 0.036$ & $8.835 \pm 0.039$ & $0.1777 \pm 0.0049$ & 0.69 & 1.31 \\   
     Blue cloud & 1948 & $1.472 \pm 0.025$ & $8.859 \pm 0.026$ & $0.1611 \pm 0.0028$ & 0.74 & 1.27 \\    
     
   \cutinhead{\em Sc galaxies}    
     All & 1843 & $1.63 \pm 0.031$ & $8.514 \pm 0.03$ & $0.2233 \pm 0.0037$ & 0.65 & 1.19 \\   
     Sparse & 1098 & $1.591 \pm 0.04$ & $8.531 \pm 0.039$ & $0.2203 \pm 0.0049$ & 0.63 & 1.17 \\   
     Rich & 243 & $1.607 \pm 0.11$ & $8.573 \pm 0.12$ & $0.2168 \pm 0.01$ & 0.76 & 1.22 \\   
     Rich  N$\ge$10 & 112 & $1.684 \pm 0.14$ & $8.478 \pm 0.15$ & $0.2059 \pm 0.013$ & 0.72 & 1.23 \\   
     Rich  BCG N$\ge$10 & 8 & $0.637 \pm 0.24$ & $9.854 \pm 0.27$ & $0.08042 \pm 0.021$ & 0.9 & 1.3 \\   
     Rich  satellite  N$\ge$10 & 104 & $1.636 \pm 0.14$ & $8.509 \pm 0.14$ & $0.2014 \pm 0.015$ & 0.72 & 1.21 \\   
     Red sequence & 409 & $1.798 \pm 0.092$ & $8.247 \pm 0.092$ & $0.2034 \pm 0.0088$ & 0.7 & 1.21 \\   
     Blue cloud & 1434 & $1.636 \pm 0.034$ & $8.538 \pm 0.032$ & $0.2194 \pm 0.0039$ & 0.64 & 1.18 \\  
\enddata
\end{deluxetable}

\begin{deluxetable}{rcccccc} 
\tablecolumns{7} 
\tablewidth{0pc} 
\tabletypesize{\scriptsize} 
\tablecaption{luminosity-size R(90) relationships\label{tab:Crelations}}
\tablehead{ 
 \colhead{} & \colhead{} &  \colhead{} & \colhead{} & \colhead{} &
  \multicolumn{2}{c}{-- Interquantile Size Range -- } \\  
  \colhead{Class} & 
  \colhead{N} &
  \colhead{Slope} &
  \colhead{Intercept} &
  \colhead{Dispersion} &
  \colhead{10\% quantile} &
  \colhead{90\% quantile \vspace{0.1cm}} \\
  \colhead{} & 
  \colhead{} &
  \colhead{$[\log(L_\odot)]$} &
   \colhead{$\left[\log(L_\odot) \over \log({\rm kpc)}\right]$ \vspace{0.1cm}} &
   \colhead{$[\log(L_\odot)]$} &
  \colhead{$[\log({\rm kpc})]$} &
   \colhead{$[\log({\rm kpc})]$}
} 
\startdata
\\
   \cutinhead{\em E galaxies}  
  All & 3163 & $1.378 \pm 0.014$ & $9.026 \pm 0.016$ & $0.1425 \pm 0.0024$ & 0.65 & 1.36 \\    
  Low Conc Red Sequence & 671 & $1.354 \pm 0.027$ & $9.022 \pm 0.033$ & $0.1641 \pm 0.0057$ & 0.54 & 1.43 \\    
High Conc Red Sequence & 743 & $1.52 \pm 0.031$ & $8.885 \pm 0.032$ & $0.1239 \pm 0.0037$ & 0.68 & 1.23 \\    
Low Conc Rich & 382 & $1.274 \pm 0.043$ & $9.132 \pm 0.055$ & $0.1509 \pm 0.0089$ & 0.66 & 1.45 \\    
High Conc Rich & 328 & $1.405 \pm 0.053$ & $9.016 \pm 0.058$ & $0.1249 \pm 0.0063$ & 0.78 & 1.28 \\    
Low Conc Sparse & 213 & $1.24 \pm 0.048$ & $9.153 \pm 0.052$ & $0.1995 \pm 0.012$ & 0.42 & 1.28 \\    
High Conc Sparse & 226 & $1.544 \pm 0.065$ & $8.848 \pm 0.063$ & $0.1435 \pm 0.0081$ & 0.63 & 1.14 \\ 
\enddata
\end{deluxetable}

\begin{deluxetable}{rcccccc} 
\tablecolumns{7} 
\tablewidth{0pc} 
\tabletypesize{\scriptsize} 
\tablecaption{luminosity-size  R(50) relationships\label{tab:relationsR(50)}}
\tablehead{ 
 \colhead{} & \colhead{} &  \colhead{} & \colhead{} & \colhead{} &
  \multicolumn{2}{c}{-- Interquantile Size Range -- } \\  
  \colhead{Class} & 
  \colhead{N} &
  \colhead{Slope} &
  \colhead{Intercept} &
  \colhead{Dispersion} &
  \colhead{10\% quantile} &
  \colhead{90\% quantile \vspace{0.1cm}} \\
  \colhead{} & 
  \colhead{} &
  \colhead{$[\log(L_\odot)]$} &
   \colhead{$\left[\log(L_\odot) \over \log({\rm kpc)}\right]$ \vspace{0.1cm}} &
   \colhead{$[\log(L_\odot)]$} &
  \colhead{$[\log({\rm kpc})]$} &
   \colhead{$[\log({\rm kpc})]$}
} 
\startdata
\\

  \cutinhead{\em Elliptical galaxies}    
     All & 3163 & $1.341 \pm 0.015$ & $9.718 \pm 0.0094$ & $0.1556 \pm 0.0026$ & 0.17 & 0.872 \\ 
     Sparse & 801 & $1.325 \pm 0.033$ & $9.708 \pm 0.018$ & $0.1738 \pm 0.0049$ & 0.085 & 0.766 \\ 
     Rich & 1514 & $1.271 \pm 0.018$ & $9.776 \pm 0.013$ & $0.1368 \pm 0.0034$ & 0.26 & 0.921 \\ 
     Rich  N$\ge$10 & 766 & $1.271 \pm 0.021$ & $9.776 \pm 0.016$ & $0.1327 \pm 0.0043$ & 0.23 & 0.963 \\ 
     Rich  BCG N$\ge$10 & 374 & $0.9983 \pm 0.041$ & $10.01 \pm 0.031$ & $0.1089 \pm 0.0042$ & 0.51 & 1. \\ 
     Rich  satellite  N$\ge$10 & 392 & $1.265 \pm 0.036$ & $9.755 \pm 0.021$ & $0.1421 \pm 0.0071$ & 0.16 & 0.802 \\ 
     Red sequence & 2908 & $1.37 \pm 0.016$ & $9.699 \pm 0.0098$ & $0.1488 \pm 0.0025$ & 0.18 & 0.877 \\ 
     Blue cloud & 255 & $1.111 \pm 0.061$ & $9.881 \pm 0.037$ & $0.2058 \pm 0.013$ & 0.082 & 0.844 \\
         
   \cutinhead{\em S0 galaxies}    
     All & 1893 & $1.17 \pm 0.021$ & $9.723 \pm 0.011$ & $0.1866 \pm 0.003$ & 0.096 & 0.768 \\ 
     Sparse & 751 & $1.154 \pm 0.026$ & $9.736 \pm 0.012$ & $0.1766 \pm 0.0043$ & 0.082 & 0.747 \\ 
     Rich & 562 & $1.184 \pm 0.035$ & $9.702 \pm 0.02$ & $0.1913 \pm 0.0067$ & 0.13 & 0.797 \\ 
     Rich  N$\ge$10 & 306 & $1.156 \pm 0.043$ & $9.687 \pm 0.022$ & $0.1828 \pm 0.0072$ & 0.13 & 0.747 \\ 
     Rich  BCG N$\ge$10 & 32 & $0.677 \pm 0.11$ & $10.13 \pm 0.063$ & $0.1037 \pm 0.014$ & 0.32 & 0.82 \\ 
     Rich  satellite  N$\ge$10 & 274 & $1.1 \pm 0.054$ & $9.692 \pm 0.026$ & $0.1795 \pm 0.007$ & 0.098 & 0.741 \\ 
     Red sequence & 1520 & $1.152 \pm 0.024$ & $9.713 \pm 0.012$ & $0.1777 \pm 0.0037$ & 0.096 & 0.753 \\ 
     Blue cloud & 373 & $1.14 \pm 0.05$ & $9.81 \pm 0.029$ & $0.2038 \pm 0.0081$ & 0.093 & 0.815 \\    
   \cutinhead{\em Sa galaxies}    
     All & 1955 & $1.101 \pm 0.021$ & $9.685 \pm 0.013$ & $0.1834 \pm 0.0034$ & 0.16 & 0.878 \\ 
     Sparse & 883 & $1.101 \pm 0.034$ & $9.677 \pm 0.021$ & $0.1755 \pm 0.0046$ & 0.16 & 0.86 \\ 
     Rich & 484 & $1.074 \pm 0.035$ & $9.719 \pm 0.024$ & $0.1796 \pm 0.006$ & 0.17 & 0.907 \\ 
     Rich  N$\ge$10 & 237 & $0.995 \pm 0.056$ & $9.724 \pm 0.036$ & $0.1911 \pm 0.0086$ & 0.17 & 0.883 \\ 
     Rich  BCG N$\ge$10 & 22 & $0.8248 \pm 0.18$ & $9.97 \pm 0.15$ & $0.1306 \pm 0.015$ & 0.46 & 0.958 \\ 
     Rich  satellite  N$\ge$10 & 215 & $0.94 \pm 0.07$ & $9.742 \pm 0.042$ & $0.192 \pm 0.0092$ & 0.16 & 0.828 \\ 
     Red sequence & 1001 & $0.9785 \pm 0.027$ & $9.732 \pm 0.015$ & $0.1896 \pm 0.0055$ & 0.14 & 0.845 \\ 
     Blue cloud & 954 & $1.219 \pm 0.029$ & $9.628 \pm 0.021$ & $0.1715 \pm 0.0039$ & 0.26 & 0.899 \\
         
   \cutinhead{\em Sb galaxies}    
     All & 2736 & $1.243 \pm 0.02$ & $9.499 \pm 0.015$ & $0.1826 \pm 0.0028$ & 0.33 & 0.955 \\ 
     Sparse & 1281 & $1.23 \pm 0.03$ & $9.499 \pm 0.022$ & $0.1799 \pm 0.0041$ & 0.33 & 0.935 \\ 
     Rich & 615 & $1.279 \pm 0.04$ & $9.496 \pm 0.032$ & $0.1831 \pm 0.0058$ & 0.36 & 0.977 \\ 
     Rich  N$\ge$10 & 264 & $1.215 \pm 0.063$ & $9.501 \pm 0.047$ & $0.1944 \pm 0.0095$ & 0.33 & 0.952 \\ 
     Rich  BCG N$\ge$10 & 36 & $0.6232 \pm 0.19$ & $10.15 \pm 0.16$ & $0.1265 \pm 0.017$ & 0.54 & 1.04 \\ 
     Rich  satellite  N$\ge$10 & 228 & $1.146 \pm 0.078$ & $9.521 \pm 0.053$ & $0.1875 \pm 0.0094$ & 0.32 & 0.925 \\ 
     Red sequence & 788 & $1.094 \pm 0.035$ & $9.563 \pm 0.025$ & $0.1991 \pm 0.0053$ & 0.24 & 0.97 \\ 
     Blue cloud & 1948 & $1.321 \pm 0.025$ & $9.458 \pm 0.018$ & $0.1722 \pm 0.0029$ & 0.38 & 0.947 \\
         
   \cutinhead{\em Sc galaxies}    
     All & 1843 & $1.525 \pm 0.032$ & $9.113 \pm 0.022$ & $0.2296 \pm 0.0039$ & 0.31 & 0.871 \\ 
     Sparse & 1098 & $1.472 \pm 0.038$ & $9.122 \pm 0.025$ & $0.2253 \pm 0.0052$ & 0.28 & 0.847 \\ 
     Rich & 243 & $1.565 \pm 0.091$ & $9.146 \pm 0.067$ & $0.2248 \pm 0.01$ & 0.41 & 0.887 \\ 
     Rich  N$\ge$10 & 112 & $1.554 \pm 0.13$ & $9.134 \pm 0.091$ & $0.2139 \pm 0.014$ & 0.38 & 0.887 \\ 
     Rich  BCG N$\ge$10 & 8 & $0.6245 \pm 0.3$ & $10.09 \pm 0.25$ & $0.08414 \pm 0.025$ & 0.55 & 0.893 \\ 
     Rich  satellite  N$\ge$10 & 104 & $1.498 \pm 0.15$ & $9.151 \pm 0.1$ & $0.2074 \pm 0.015$ & 0.38 & 0.884 \\ 
     Red sequence & 409 & $1.661 \pm 0.078$ & $8.929 \pm 0.052$ & $0.2133 \pm 0.007$ & 0.36 & 0.893 \\ 
     Blue cloud & 1434 & $1.53 \pm 0.037$ & $9.138 \pm 0.025$ & $0.2261 \pm 0.0051$ & 0.3 & 0.865 \\
     \enddata
\end{deluxetable}

\begin{deluxetable}{rcccccc} 
\tablecolumns{7} 
\tablewidth{0pc} 
\tabletypesize{\scriptsize} 
\tablecaption{Sersic luminosity- Effective Radii relationships\label{tab:Reffrelations}}
\tablehead{ 
 \colhead{} & \colhead{} &  \colhead{} & \colhead{} & \colhead{} &
  \multicolumn{2}{c}{-- Interquantile Size Range -- } \\  
  \colhead{Class} & 
  \colhead{N} &
  \colhead{Slope} &
  \colhead{Intercept} &
  \colhead{Dispersion} &
  \colhead{10\% quantile} &
  \colhead{90\% quantile \vspace{0.1cm}} \\
  \colhead{} & 
  \colhead{} &
  \colhead{$[\log(L_\odot)]$} &
   \colhead{$\left[\log(L_\odot) \over \log({\rm kpc)}\right]$ \vspace{0.1cm}} &
   \colhead{$[\log(L_\odot)]$} &
  \colhead{$[\log({\rm kpc})]$} &
   \colhead{$[\log({\rm kpc})]$}
} 
\startdata
\\
\cutinhead {\em Elliptical galaxies}     
    All & 3156 & $1.379  \pm 0.015$ & $9.639  \pm 0.011$ & $0.1453 \pm 0.0027$ & 0.24 & 0.945  \\  
    Sparse & 800 & $1.364  \pm 0.03$ & $9.634  \pm 0.018$ & $0.1646  \pm 0.0047$ & 0.14 & 0.838  \\  
    Rich & 1509 & $1.311  \pm 0.022$ & $9.698  \pm 0.017$ & $0.1281  \pm 0.0031$ & 0.33 & 0.985  \\  
    Rich  N$\ge$10 & 764 & $1.304  \pm 0.022$ & $9.707  \pm 0.017 $ & $0.1238  \pm 0.0042$ & 0.28 & 1.01  \\  
    Rich  BCG N$\ge$10 & 374 & $1.098  \pm 0.04$ & $9.9  \pm 0.034 $ & $0.1026  \pm 0.0035$ & 0.58 & 1.05  \\  
    Rich  satellite  N$\ge$10 & 390 & $1.266  \pm 0.036$ & $9.707  \pm 0.023$ & $0.1347  \pm 0.0065$ & 0.23 & 0.875  \\  
    Red sequence & 2904 & $1.403  \pm 0.013$ & $9.62  \pm 0.0093$ & $0.1386  \pm 0.0025$ & 0.25 & 0.947  \\  
    Blue cloud & 252 & $1.176  \pm 0.06$ & $9.803  \pm 0.041$ & $0.1963  \pm 0.013$ & 0.14 & 0.893  \\    
  \cutinhead {\em S0 galaxies}
    All & 1887 & $1.24  \pm 0.019$ & $9.662  \pm 0.01$ & $0.1791  \pm 0.0037$ & 0.15 & 0.807  \\  
    Sparse & 749 & $1.195  \pm 0.037$ & $9.69  \pm 0.018$ & $0.172  \pm 0.0057$ & 0.13 & 0.777  \\  
    Rich & 561 & $1.278  \pm 0.033$ & $9.629  \pm 0.019$ & $0.1819  \pm 0.0054$ & 0.19 & 0.826  \\  
    Rich  N$\ge$10 & 306 & $1.224  \pm 0.046$ & $9.631  \pm 0.027 $& $0.1759  \pm 0.007$ & 0.17 & 0.77  \\  
    Rich  BCG N$\ge$10 & 32 & $0.7051  \pm 0.12$ & $10.11  \pm 0.07$ & $0.1036  \pm 0.011$ & 0.36 & 0.855  \\  
    Rich  satellite  N$\ge$10 & 274 & $1.172  \pm 0.052$ & $9.638  \pm 0.025$ & $0.173  \pm 0.0068$ & 0.16 & 0.764  \\  
    Red sequence & 1518 & $1.217  \pm 0.026$ & $9.656  \pm 0.013$ & $0.1711  \pm 0.0042$ & 0.16 & 0.78  \\  
    Blue cloud & 369 & $1.231  \pm 0.05$ & $9.738  \pm 0.03$ & $0.1931  \pm 0.0079$ & 0.14 & 0.844  \\    
  \cutinhead {\em Sa galaxies}
    All & 1952 & $1.189  \pm 0.019$ & $9.627  \pm 0.013$ & $0.1776  \pm 0.0038$ & 0.22 & 0.889  \\  
    Sparse & 882 & $1.194  \pm 0.032$ & $9.618  \pm 0.02$ & $0.1687 \pm 0.0048$ & 0.22 & 0.871  \\  
    Rich & 483 & $1.144  \pm 0.041$ & $9.669  \pm 0.028$ & $0.1765  \pm 0.0063$ & 0.22 & 0.911  \\  
    Rich  N$\ge$10 & 237 & $1.054  \pm 0.064$ & $9.682  \pm 0.039 $& $0.1884  \pm 0.0085$ & 0.22 & 0.901  \\  
    Rich  BCG N$\ge$10 & 22 & $0.857  \pm 0.21$ & $9.948  \pm 0.17 $ & $0.1316  \pm 0.015$ & 0.51 & 0.959  \\  
    Rich  satellite  N$\ge$10 & 215 & $1.002  \pm 0.07$ & $9.7  \pm 0.044$ & $0.1893  \pm 0.0094$ & 0.22 & 0.862  \\  
    Red sequence & 999 & $1.067  \pm 0.032$ & $9.675  \pm 0.02$ & $0.1834  \pm 0.0048$ & 0.2 & 0.87  \\  
    Blue cloud & 953 & $1.296  \pm 0.033$ & $9.578  \pm 0.023$ & $0.1661  \pm 0.0048$ & 0.3 & 0.903  \\    
  \cutinhead {\em Sb galaxies}
    All & 2733 & $1.314  \pm 0.02$ & $9.452  \pm 0.015$ & $0.1795  \pm 0.0028$ & 0.37 & 0.958  \\  
    Sparse & 1279 & $1.303  \pm 0.031$ & $9.453  \pm 0.023$ & $0.1763  \pm 0.0039$ & 0.35 & 0.938  \\  
    Rich & 614 & $1.355  \pm 0.041$ & $9.444  \pm 0.032$ & $0.1796  \pm 0.0068$ & 0.4 & 0.99  \\  
    Rich  N$\ge$10 & 263 & $1.311  \pm 0.066$ & $9.437  \pm 0.053 $ & $0.1893  \pm 0.01$ & 0.39 & 0.952  \\  
    Rich  BCG N$\ge$10 & 36 & $0.6799  \pm 0.16$ & $10.11  \pm 0.14$ & $0.1251  \pm 0.015$ & 0.59 & 1.04  \\  
    Rich  satellite  N$\ge$10 & 227 & $1.245  \pm 0.077$ & $9.458  \pm 0.054$ & $0.1838  \pm 0.01$ & 0.38 & 0.926  \\  
    Red sequence & 788 & $1.185  \pm 0.034$ & $9.501  \pm 0.025$ & $0.1941  \pm 0.0053$ & 0.28 & 0.974  \\  
    Blue cloud & 1945 & $1.374  \pm 0.023$ & $9.426  \pm 0.018$ & $0.1699  \pm 0.0031$ & 0.4 & 0.95  \\    
  \cutinhead \em Sc galaxies     
    All & 1843 & $1.561  \pm 0.03$ & $9.093  \pm 0.02$ & $0.2285  \pm 0.0042$ & 0.32 & 0.874  \\  
    Sparse & 1098 & $1.509  \pm 0.037$ & $9.102  \pm 0.025$ & $0.2242  \pm 0.0051$ & 0.28 & 0.848  \\  
    Rich & 243 & $1.606  \pm 0.095$ & $9.12  \pm 0.067$ & $0.2238  \pm 0.0096$ & 0.43 & 0.889  \\  
    Rich  N$\ge$10 & 112 & $1.593  \pm 0.14$ & $9.11  \pm 0.1$ & $0.2125  \pm 0.015$ & 0.39 & 0.888  \\  
    Rich  BCG N$\ge$10 & 8 & $0.6577  \pm 0.28$ & $10.07  \pm 0.23 $ & $0.08345  \pm 0.023$ & 0.56 & 0.904  \\  
    Rich  satellite  N$\ge$10 & 104 & $1.538  \pm 0.14$ & $9.127  \pm 0.099$ & $0.206  \pm 0.014$ & 0.39 & 0.886  \\  
    Red sequence & 409 & $1.704  \pm 0.079$ & $8.902  \pm 0.054$ & $0.2111  \pm 0.0081$ & 0.36 & 0.893  \\  
    Blue cloud & 1434 & $1.567  \pm 0.035$ & $9.118  \pm 0.023$ & $0.2249  \pm 0.0046$ & 0.3 & 0.867  \\   
    \enddata
\end{deluxetable}

\begin{deluxetable}{rcccccc} 
\tablecolumns{7} 
\tablewidth{0pc} 
\tabletypesize{\scriptsize} 
\tablecaption{de Vaucouleurs luminosity-size relationships\label{tab:Devrelations}}
\tablehead{ 
 \colhead{} & \colhead{} &  \colhead{} & \colhead{} & \colhead{} &
  \multicolumn{2}{c}{-- Interquantile Size Range -- } \\  
  \colhead{Class} & 
  \colhead{N} &
  \colhead{Slope} &
  \colhead{Intercept} &
  \colhead{Dispersion} &
  \colhead{10\% quantile} &
  \colhead{90\% quantile \vspace{0.1cm}} \\
  \colhead{} & 
  \colhead{} &
  \colhead{$[\log(L_\odot)]$} &
   \colhead{$\left[\log(L_\odot) \over \log({\rm kpc)}\right]$ \vspace{0.1cm}} &
   \colhead{$[\log(L_\odot)]$} &
  \colhead{$[\log({\rm kpc})]$} &
   \colhead{$[\log({\rm kpc})]$}
} 
\startdata
\\
   \cutinhead{\em E galaxies}  
 All & 3162 & $1.1  \pm 0.014$ & $9.734  \pm 0.011$ & $0.1821  \pm 0.0036$ & 0.3 & 1.15  \\    
Sparse & 801 & $1.058  \pm 0.033$ & $9.727  \pm 0.022$ & $0.2004  \pm 0.0061$ & 0.21 & 1.02  \\    
Rich & 1514 & $1.054  \pm 0.015$ & $9.8  \pm 0.014$ & $0.158  \pm 0.0049$ & 0.41 & 1.24  \\    
Rich  N$\ge$10 & 766 & $1.064  \pm 0.022$ & $9.789  \pm 0.02$ & $0.1553  \pm 0.0062$ & 0.39 & 1.31  \\    
Rich  BCG N$\ge$10 & 374 & $0.8154  \pm 0.03$ & $10.08  \pm 0.031$ & $0.1186  \pm 0.0046$ & 0.67 & 1.41  \\    
Rich  satellite  N$\ge$10 & 392 & $1.042  \pm 0.037$ & $9.765  \pm 0.027$ & $0.1659  \pm 0.0098$ & 0.27 & 1.03  \\    
Red sequence & 2907 & $1.137  \pm 0.013$ & $9.704  \pm 0.011$ & $0.1747  \pm 0.0037$ & 0.32 & 1.15  \\    
Blue cloud & 255 & $0.8198  \pm 0.06$ & $9.97  \pm 0.047$ & $0.2315  \pm 0.013$ & 0.19 & 1.13  \\
\enddata
\end{deluxetable}

\end{document}